\documentclass[twoside,a4paper]{article}
\usepackage{helvet,times}
\usepackage{bm,textcomp}
\usepackage{url}
\usepackage{subcaption}
\usepackage{tikz}
\usepackage{floatrow}
\usepackage{lipsum}
\usepackage{wrapfig}
\usepackage[innercaption]{sidecap}    
\sidecaptionvpos{figure}{t}
\usepackage{xr}
\usepackage{geometry}
\newgeometry{vmargin={15mm}, hmargin={19mm,19mm}}
\usepackage{hyperref}
\makeatletter
\newcommand*{\addFileDependency}[1]{
  \typeout{(#1)}
  \@addtofilelist{#1}
  \IfFileExists{#1}{}{\typeout{No file #1.}}
}
\makeatother

\usepackage{amsmath}

\usepackage[round,numbers,sort&compress]{natbib}

\DeclareUnicodeCharacter{00A0}{ }

\DeclareUnicodeCharacter{3B1}{\ensuremath{\alpha}}

\begin{document}

\setcounter{page}{1} 

{\fontsize{17}{12}\selectfont
Stochastic model of T Cell repolarization during target elimination (II)
}\\

\begin{center}
{\fontsize{12}{12}\selectfont
Ivan Hornak$^{1,*}$, Heiko Rieger$^{1}$\\
\vspace{1mm}
$^{1}$Department of Theoretical Physics, Center for Biophysics, Saarland University,
Saarbr{\"u}cken, Germany\\
\vspace{1mm}
$^{*}$hornak@lusi.uni-sb.de
}\\
\end{center}

%
%

\section*{Abstract}
Cytotoxic T lymphocytes (T cells) and natural killer cells form a tight
contact, the immunological synapse (IS), with target cells, where they
release their lytic granules containing perforin/granzyme and cytokine
containing vesicles. During this process the cell repolarizes and moves the microtubule organizing center (MTOC) towards the IS.  In the first part of our work
 we developed a
computational model for the molecular-motor-driven motion of the MT
cytoskeleton during T cell
polarization and analyzed effects of cortical sliding and
capture-shrinkage mechanisms. Here we use this model to analyze the dynamics of the MTOC
repositioning in situations in which a) the IS is in
an arbitrary position with respect to the initial position of the MTOC
and b) the T cell has two IS at two arbitrary positions.
In the case of one IS, we found that the initial position determines which mechanism is dominant and that the time of repositioning does not rise monotonously with the MTOC-IS distance.
In the case of two IS,
we observe
several scenarios that have also been reported experimentally: the MTOC
alternates stochastically (but with a well defined average transition
time) between the two IS; it wiggles in between the two IS without
transiting to one of the two; or it is at some point pulled to one of
the two IS and stays there. Our model allows to predict which scenario
emerges in dependency of the mechanisms in action and the number of
dyneins present.
We report that
 the presence of capture-shrinkage mechanism at least in one IS is necessary to assure the transitions
 in every cell configuration. 
Moreover, the frequency of transitions does not decrease with the distance between the two IS and is the highest when both mechanisms are present in both IS.

\section*{Introduction}
T Cells have a key role in the adaptive branch of our immune system by finding and destruction of virus-infected and tumor cells, parasites, and foreign invaders. Cytotoxic killing of a target cell is achieved in three subsequent steps. First, T Cell binds to the surface of the target cell and creates a tight contact zone called immunological  synapse (IS)
\cite{rudolph_how_2006,garcia_reconciling_2012, zinkernagel_restriction_1974,attaf_t_2015,wucherpfennig_t_2004,
babbitt_binding_1985,monks_three-dimensional_1998,dustin_novel_1998,dustin_understanding_2010}. Second, the T Cell relocates the microtubule organizing center (MTOC) towards the IS by a massive movement of the entire MT cytoskeleton due to forces acting on MTs
\cite{geiger_spatial_1982,kupfer_polarization_1982,yi_centrosome_2013,
stinchcombe_centrosome_2006,maccari_cytoskeleton_2016,kuhn_dynamic_2002,
hui_dynamic_2017}.
This process involves the repositioning of mitochondria, the Golgi apparatus, and the endoplasmic reticulum, since the organelles are bound to the cytoskeleton and relocate with it  \cite{maccari_cytoskeleton_2016,kupfer_reorientation_1984,kupfer_specific_1986,gurel_connecting_2014,lee_dynamic_1988,waterman-storer_endoplasmic_1998,palmer_role_2005}. 
In the third step, the T Cell releases the cytotoxic material from the lytic granules towards the target cell leading to its death by necrosis or apthosis \cite{mullbacher_granzymes_1999,lowin_perforin_1995,voskoboinik_perforin-mediated_2006,grossman_orphan_2003,krzewski_human_2012,
groscurth_killing_1998}. 
The secretion of lytic granules can take place 
without the MTOC repolarization \cite{golstein_early_2018}, or before it \cite{bertrand_initial_2013}.  
However, it does not make the repositioning redundant, since the MTOC accompanied granule secretion may be crucial for the killing of 
 resistant cells.
\newline

 The IS is divided into several supramolecular activation clusters (SMACs) including ring shaped  peripheral SMAC (pSMAC)
\cite{monks_three-dimensional_1998,dustin_understanding_2010,andre_use_1990,lin_c-smac_2005,choudhuri_signaling_2010}.
Dynein, a minus-end-directed (toward the MTOC) molecular motor protein, is indispensable for the repositioning as was shown by knock out experiments
\cite{martin-cofreces_mtoc_2008,nguyen-ngoc_coupling_2007,saito_mcp5_2006,yamashita_fission_2006,
ananthanarayanan_dynein_2013}. The dynein colocalizes with the adaptor protein ADAP that forms a ring at the IS periphery after the activation of the T cell \cite{combs_recruitment_2006,hashimoto-tane_dynein-driven_2011}. Dynein plays a key role in the two mechanisms proposed to drive the repositioning: cortical sliding and capture-shrinkage. 
In the cortical sliding mechanism the dyneins step to the minus-end of MTs(towards the MTOC) while being anchored on the cell membrane and therefore pull the MTOC towards the IS \cite{combs_recruitment_2006,stinchcombe_communication_2014,kuhn_dynamic_2002}. It was indicated that the ring shaped  pSMAC is the place where attached dyneins are anchored \cite{combs_recruitment_2006,kuhn_dynamic_2002}. 
It was shown in \cite{sanchez_actin_2019} that the recruitment 
of the dynein to the IS is correlated 
and promoted by
the depletion of cortical actin filaments from the same place.
\newline

A detailed analysis of the capture-shrinkage process was performed by Yi et al 
\cite{yi_centrosome_2013}.
An optical trap was used to place the target cell so that the IS(contact zone) is initially diametrically opposed to the MTOC. This well defined initial configuration allowed quantitative dynamical imagining including the observation of the MT cytoskeleton morphology. They provided  strong evidence that the repositioning is driven by a capture-shrinkage mechanism \cite{laan_cortical_2012} involving the depolymerization
of the caught MT in a confined area in the center of the IS. It was shown \cite{yi_centrosome_2013}  that  MTs bend alongside the cell membrane to reach the IS. Consequently, the MTs caught by their plus-end in the center of the IS straighten, form a narrow stalk and depolymerize at the capture-point. The MTOC is pulled to the center of the IS which invaginates the cell indicating the location of the main pulling force. The capture shrinkage mechanism was identified as the main driving force of the repositioning, since inhibiting the MT depolymerization substantially slowed down the repositioning. Yi et al. \cite{yi_centrosome_2013} reported that the repositioning can be divided into two phases that differ in the MTOC speed and the direction of its motion. In the first so-called polarization phase, the MTOC travels quickly in a circular motion around the nucleus. In the second, docking phase, the MTOC moves slowly and directly towards the IS.
 \newline

T cells can attack two target cells simultaneously, in which case two IS are established \cite{kuhn_dynamic_2002}.
In this case, the MTOC transits repeatedly between the two IS \cite{kuhn_dynamic_2002}, reminiscent of mitotic spindle oscillations in C. elegans
\cite{grill_theory_2005,kozlowski_cortical_2007,grill_spindle_2005,
kruse_oscillations_2005,colombo_translation_2003,grill_distribution_2003,
nguyen-ngoc_coupling_2007}. 
These spontaneous spindle oscillations have been explained by the cooperative attachment and detachment of cortical force generators to astral microtubules 
\cite{grill_polarity_2001,siller_spindle_2009,
couwenbergs_heterotrimeric_2007,pecreaux_spindle_2006}.
For MTOC oscillations in T cells with two synapses a similar scenario has been proposed in connection with cortical sliding mechanism 
\cite{kim_deterministic_2009}. There it has been hypothesized that MTs on the trailing side of the MTOC are lifted off the pulling surface by viscous drag in the cytoplasm leading to their detachment from cortical motors. Here we propose a different mechanism, which also works for capture-shrinkage as reported in \cite{yi_centrosome_2013} and which relies on dynamic MTs, similar to, for instance, meiotic nuclear oscillations in S.pombe
\cite{vogel_self-organization_2009,ding_oscillatory_1998,yamamoto_cytoplasmic_1999}.
\newline

The interplay between dyneins and filaments is influenced by dynamic MTs which constantly grow and shrink: periods of grow alternate with the periods of rapid depolymerization in a process called dynamic instability, DI, \citep{
walker_dynamic_1988,mitchison_dynamic_1984,vorobev_dynamics_2003,
horio_role_2014,brouhard_dynamic_2015,
mandelkow_microtubule_1991,bieling_reconstitution_2007,
desai_microtubule_1997,kerssemakers_assembly_2006,schek_microtubule_2007,gardner_microtubule_2008}. 
This process allows the cytoskeleton to adopt itself to the needs and the functions of the cell to perform substantial shape changes through the cell cycle 
\cite{horio_role_2014,myers_distinct_2011,lacroix_microtubule_2018,
fuesler_dynamic_2012}. 
Transitions between two IS would not be possible without the DI of MTs: at the end of the repositioning process towards one IS the MT cytokeleton is deformed and capture-shrinkage MTs are depolymerized \cite{yi_centrosome_2013,hornak_stochastic_2020}. Due to DI, the depolymerized MTs regrow and the deformed cytoskeleton can restructure.
\newline

The main reason for MTOC oscillations in T cells with two IS as well as mitotic spindle oscillations to occur is that one has two distinct locations of dynein accumulation on the cell boundary, where motors can catch the MTs. The attachment of dynein is stochastic and one location wins the tug-of-war between the attached dyneins, resulting in the MTOC relocation towards that location. In both cases, dyneins detach as the MTOC approaches. In the case of spindle oscillation two restoring forces pushing the spindle back to the center were considered: the cortical pushing of the MTs polymerization-driven growth against a barrier followed by the bending of the filaments, and the pulling force of the dyneins opposite to the movement of the MTOC 
\cite{wu_forces_2017,garzon-coral_force-generating_2016,howard_elastic_2006,
pecreaux_mitotic_2016,
kozlowski_cortical_2007,dogterom_measurement_1997,
grill_theory_2005,grill_spindle_2005}.  
The purpose of our theoretical study is to elucidate the potential mechanisms for cooperative dynein attachment and 
detachment leading to stable, bi-stable or oscillatory MTOC relocation in T cells with two IS.
\newline

\section*{Computational model}
\label{computational_model}
\begin{figure}[hbt!]
\centering
     \includegraphics[trim=0 490 15 10,clip,width=0.8\textwidth]{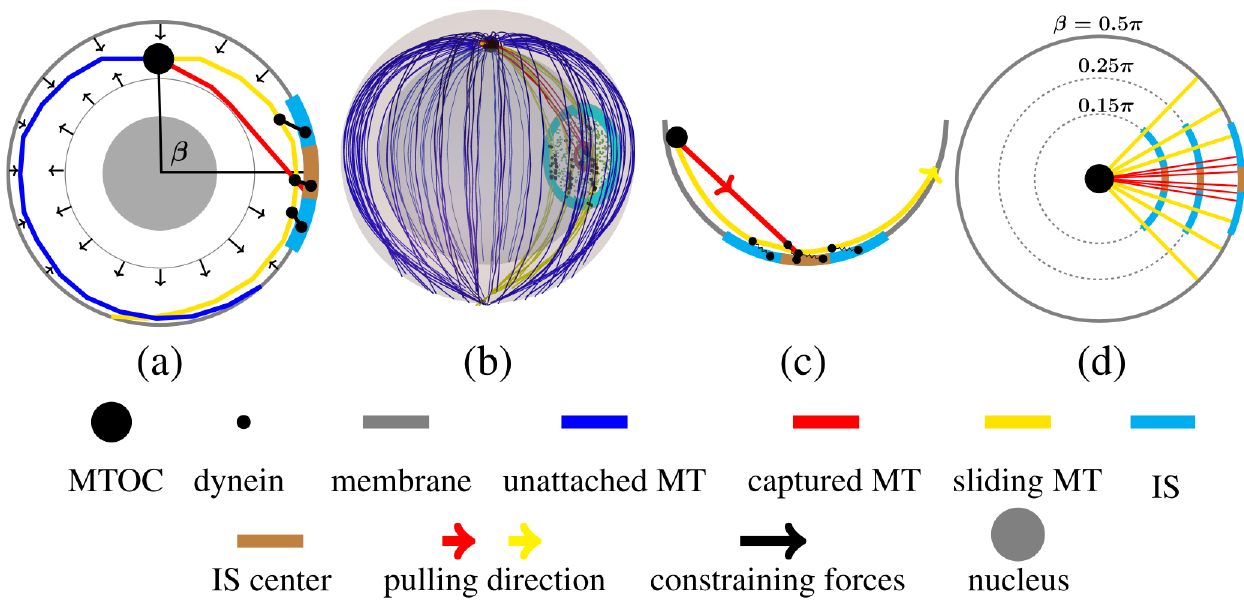}    
\caption{
{\small
 (a–d) Sketch of the model of the cell with one IS. 
 (a) A two-dimensional cross-section of the model is shown.
The movement of the MTs sprouting from the MTOC is confined between the nucleus and the cell membrane.
MTs are pulled by the capture shrinkage and cortical sliding mechanisms employing dynein motors in the IS.
The configuration of the cell is determined by the angle $\beta$ between the IS and the initial position of the MTOC. 
(b)  A three-dimensional sketch of the cell model is given. 
The plasma membrane and the nucleus are represented by the transparent
outer and inner spheres, respectively. 
Small green spheres represent unattached dyneins in the IS and its center depicted by the cyan and small brown disks, respectively.
Cortical sliding dyneins are anchored in the IS and the capture-shrinkage dyneins in the center. 
(c) A sketch of the cortical sliding mechanism and the capture-shrinkage mechanism is shown.
Small black dots on the membrane and on the MTs represent dynein's anchor and attachment points, respectively.
When the  MT is pulled by capture-shrinkage mechanism, the end of the MT is anchored in the center of the IS and depolymerizes.
Cortical sliding MTs slide on the surface and the plus-end remains free.
(d) Sketch of MTs intersecting the IS and its center in the cells with different angles $\beta$.
The percentage of MTs intersecting the IS is given by the ratio of the 
diameter of the IS to the cell circumference corresponding to angle $\beta$ depicted by dashed circles.
The percentage decreases to the minimum at $\beta=0.5\pi$ and then it increases again. 
  \label{fig:variable_Beta_basic} 
}  
}
\end{figure}

{\small
We use the computational model introduced in \cite{hornak_stochastic_2020}.
The cell membrane and the nucleus are represented by  two spheres with radius 5$\mu$m  and 3.8$\mu$m, respectively.
MTs sprout from the MTOC to the cell periphery, as sketched in Figs. 
\ref{fig:variable_Beta_basic}a and b.
They
are modeled  by a bead-rod model with constrained Langevin dynamics.
The MTs move under the influence of several forces: bending, drag, molecular motors, noise and repulsive forces keeping them between the nucleus and the cell membrane.
The MTOC moves to the IS due to the pulling force of dyneins acting via two mechanisms: cortical sliding during which the plus-end of the MT remains free and filament slides tangentially along the plasma membrane, and capture-shrinkage, by which dyneins capture the tip of the MT and depolymerize it by pulling it against the membrane, 
as sketched in Fig. \ref{fig:variable_Beta_basic}c.
Dyneins acting via cortical sliding and capture-shrinkage are located in the complete IS and the narrow center, respectively.
The two regions are
represented by intersections of the cell sphere with cylinders with radius 
$R_{\rm IS}=2\mu\textrm{m}$ for the complete IS and
$R_{\rm CIS}=0.4\mu\textrm{m}$ 
for the center, as sketched in Fig.  \ref{fig:variable_Beta_basic}.
Note that we assume the dyneins to be immobile and firmly fixed at the cell boundary. This is in contrast to a recently proposed model in which dyneins can more or less freely (dependent upon assumed friction coefficients) move in an actin depleted zone of the IS \cite{gros_dynein_2021} and thus self-organizing into clusters at the boundary of the actin depleted zone. This is an interesting hypothesis that has still to be confirmed or refuted by experiments, but we do not expect dyneins that can move within a restricted region of the IS to alter our main conclusions substantially, for which reason we stick to our original model of a pre-defiend dynein location \cite{hornak_stochastic_2020}.
In \cite{hornak_stochastic_2020} we focused on the analysis of the MTOC repositioning process in the experimental setup used in \cite{yi_centrosome_2013}, in which the MTOC and the IS are initially diametrically opposed.
Here we consider naturally occurring situations, in which the angle $\beta$ between the MTOC and the IS (see \ref{fig:variable_Beta_basic}a) is arbitrary, and situations in which the T cell attaches simultaneously to two target cells and thus forms two IS \cite{kuhn_dynamic_2002}.
}
\newline

\begin{figure}[hbt!]
\centering
     \includegraphics[trim=0 410 0 0,clip,width=\textwidth]{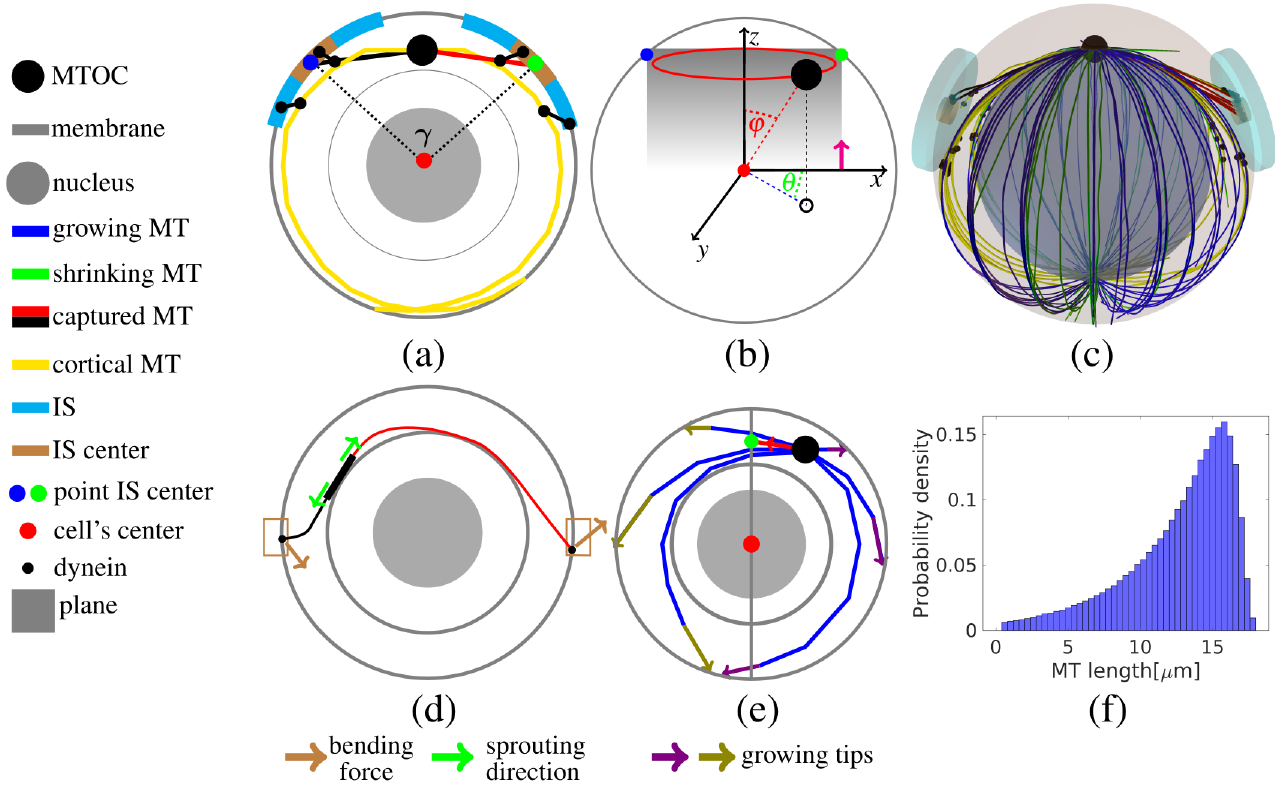}    
   \caption{ {\small
   Sketch of the model of the cell with two IS.
(a) The sketch of the angle $\gamma$ between two IS .
Black dotted lines depict the axes of both IS with the mutual angle $\gamma$.
Small black dots on the membrane and MTs represent dynein 
anchor and attachment points, respectively.
MTs can attach to capture-shrinkage or cortical sliding dyneins in both IS.
(b) The sketch of the azimuthal and polar angles is given.
The positions of the centers of both IS and the cell are located on the 
$xz$ plane 
depicted in gray.
When $\gamma<\pi$ both IS are located in the upper hemisphere($z>0$) denoted by the
magenta arrow.
The polar angle $\varphi$ describes the cone with the vertex located in the center of the cell. 
The azimuthal angle $\theta$ gives the angle between the projection 
of the MTOC position on $xy$ plane depicted by the small black circle and x-Axis.
(c) 
 Three-dimensional sketch of the cell
model is given, 
$\gamma = \frac{3}{4} \pi$.
Growing and shrinking MTs sprout from the MTOC to the cell periphery and
can be pulled by the two mechanisms in both IS. Dyneins
in one IS cooperate and dyneins from different IS are in a tug-of-war.
(d) The sketch of bending forces acting on MTs attached to the capture-shrinkage dynein is given.
The wide black line and brown rectangles represent the plane of the MTOC and the centers of IS, respectively.
Attached MTs sprout from the MTOC tangentially.
Bending forces push the long MT against the cell membrane and pull the short MT from it.
(e)
 A two dimensional sketch of the forces acting on the cytoskeleton.
The gray line represents the $xz$ plane on which the centers of the cell
and both IS are located. 
The red line stands for the MT attached in the IS.
Dynein forces acting on the red MT pull the MTOC to the $xz$ plane where the IS is located.
The growing olive and violet plus-end of MTs push the MTOC from the $xz$ plane and towards it,
respectively.
(f) The probability density of MTs length.
\label{fig:two_IS_sketch} } }  
 \end{figure}

{\small
To analyze the situation with two IS we augmented our model presented in
\cite{hornak_stochastic_2020}
in several ways.
The configuration of a cell with two IS is defined by the angle
$\gamma$ between the lines connecting the centers of IS with the 
center of the cell, sketched in Fig. \ref{fig:two_IS_sketch}a.
Both IS and the center of the cell are located on the $xz$ plane of the coordinate system, sketched in Fig.  \ref{fig:two_IS_sketch}b and visually demonstrated in Fig.  \ref{fig:two_IS_sketch}c.
The dyneins from both IS are in a tug-of-war leading to an increase of the detachment rate \cite{hornak_stochastic_2020}.
When all capture-shrinkage dyneins detach from the MT,
the plus end is no longer fixed on the cell membrane. 
Most importantly we
included the dynamical instability of MTs
 \citep{mitchison_dynamic_1984,desai_microtubule_1997,brouhard_dynamic_2015,
horio_role_2014,zhang_mechanistic_2015}
since we hypothesized that transitions between two IS rely on DI.
The measured values of parameters of dynamic instability differ \cite{cassimeris_real-time_1988,sammak_direct_1988,belmont_real-time_1990,mitchison_dynamic_1984,zwetsloot_measuring_2018,
steinberg_microtubules_2001,yvon_non-centrosomal_1997,carminati_microtubules_1997,adames_microtubule_2000,
drummond_dynamics_2000,van_damme_vivo_2004},
 they depend on the cell phase \cite{yamashita_three-dimensional_2015,tirnauer_yeast_1999}, and on the distance from the cell membrane \cite{komarova_life_2002,alieva_microtubules_2010,brunner_clip170-like_2000,rusan_cell_2001}.
We take the following estimates from the literature: growth velocity $v_g=0.1 \mu\textrm{ms}^{-1}$ \citep{brouhard_dynamic_2015,zwetsloot_measuring_2018,trushko_growth_2013,
van_damme_vivo_2004} - although it might depend mildly on load and MT plus end location \cite{alieva_microtubules_2010,schek_microtubule_2007}; shrinking velocity $v_s=0.2\mu\textrm{ms}^{-1}$; rescue rate (the transition rate from shrinkage to growth) 
$r_r=0.044\textrm{s}^{-1}$ \citep{cassimeris_real-time_1988,
 shelden_observation_1993,
 belmont_real-time_1990,
 fees_unified_2019}; and a length dependent catastrophe rate (transition rate from growth to shrinkage) 
$c_{r}(L) = \textrm{exp}(( L - L_{c} ) / b_{c})\textrm{s}^{-1}$
where $L_{c} = \pi R_{\textrm{Cell}} + \frac{ R_{\textrm{Cell}} }{2}$,
$b_{c} = ( L_{0} - L_{c} ) / \textrm{ln}( r_{c} )$, 
$L_{0} = \pi R_{\textrm{Cell}}$ and $r_{c} = 0.022$, 
  reflecting a lower catastrophe rate close to the MTOC and a higher one at the cell periphery \cite{tischer_force-_2009,komarova_life_2002,myers_distinct_2011}. 
The MT length distribution resulting from the
dynamic instability with aforementioned parameters is shown in Fig.
\ref{fig:two_IS_sketch}f.
Due to the dynamic instability, growing and shrinking MTs coexist in a dynamically changing cytoskeleton affected by the two mechanisms in both IS, as visualized in Fig. \ref{fig:two_IS_sketch}c.
The dynamic instability adds another force acting on MTs, since the growing tips of filaments are pushed against the cell membrane, sketched in Fig. \ref{fig:two_IS_sketch}e.
In contrast to the dynein forces, growing tips 
can push the MTOC from both IS and the $xz$ plane.
Since the plus-end of the MTs remain free during the cortical sliding mechanism, the MTs can grow or shrink even when attached.
The MT length influences the contact between MTs and the dyneins  on the cell membrane.
The MTOC is modeled as a planar structure \cite{hornak_stochastic_2020}  and the MT sprouts from the MTOC radially. 
The short MTs have to bend to stay in contact with the dyneins on the membrane, as  sketched in Fig. \ref{fig:two_IS_sketch}d.
Once the dynein detach, the tip recedes from the membrane making the reattachment unlikely.
On the other hand, bending forces press the tip of a long MT against the cell membrane where it can attach to dyneins.}
\newline

{\small
We implemented the model in C++ on a computer cluster with  Intel(R) Xeon(R) CPU E5-2660 0 @ 2.20GH processors,
 Linux operating system (Arch-Linux 4.1.7-hardened-r1) and
compiler  g++ 4.9.2 and 
  performed simulation runs to generate the data shown in this publication. 
 The program listing is publicly available on GitHub  \cite{noauthor_ihornak_nodate}. 
 The snapshots from the simulations and supplementary movies were made in POV-Ray.}

\section*{Results}

\subsection*{Repositioning time scales}
Before we present the results of computer simulations of the model defined in \ref{computational_model} 
the previous section,
we give an estimate for the time scale of the MTOC repositioning process based on the antagonistic interplay of friction and pulling forces acting on the MT cytoskeleton and compare it with the  repositioning times from experiments.
\newline

\begin{figure}[hbt!]
\centering
     \includegraphics[trim=0 650 15 0,clip,width=\textwidth]{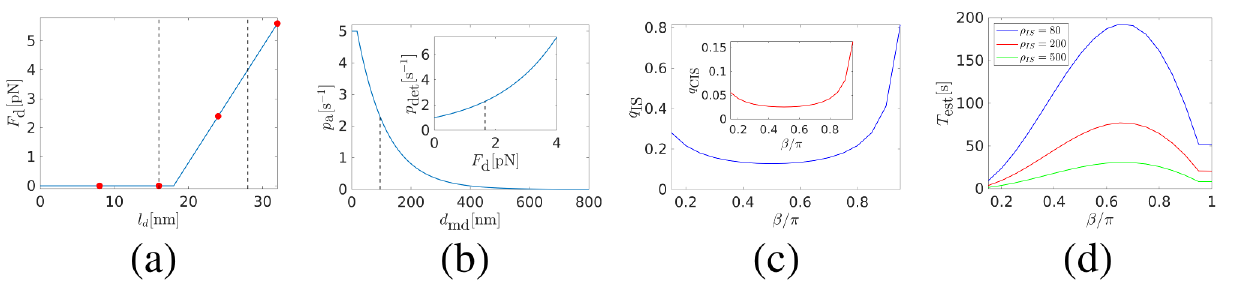} 
\caption{ Estimations of dynein forces in IS.  
 (a) Dependence of the dynein force $F_{d}$ on the length of the dynein stalk $l_{d}$.
The red points represent the multiples of the  dynein step
$d_{\textrm{step}} = 8\textrm{nm}$.
Dynein makes the first two steps quickly due to the zero load and
the stepping slows down as the force increases and stops at the stall force $F_{S}=4\textrm{pN}$.
The dashed black lines represent  the second step and the length corresponding to the stall force and delimit the probable length of the 
dynein stalk.
(b)  The dependence of the dynein attachment rate $p_{\textrm{a}}$ on the distance $d_{\textrm{md}}$ between the dynein and the MT.
The dependence of the dynein detachment rate $p_{\textrm{det}}$  on
the dynein force $F_{d}$ in the inset.
The black dashed curve in the inset represents the detachment rate
$p_{\textrm{det}}(\overline{F_{\textrm{d}}})$ corresponding to the average dynein force.
The black dashed line in the main picture represent 
the distance $d_{\textrm{md}}$ when $p_{\textrm{a}} = p_{\textrm{det}}(\overline{F_{\textrm{d}}})$.
(c) Fraction of all MTs intersecting the IS $q_{\textrm{IS}}$(inset: central area of the IS $q_{\textrm{CIS}}$) as function of the angle $\beta$. 
(d) The dependence of the estimated time of the repositioning $T_{\textrm{est}}$
on the angle $\beta$. 
Comparison of the Figs. d and 
\ref{fig:variable_Beta}b shows that the estimated times of
the repositioning are comparable to the measured ones and that the dependencies on the angle $\beta$ have the same development, see 
Fig. S2l.
The estimated times are shorter when $\rho_{\textrm{IS}}=200, 500\mu\textrm{m}^{-2}$, since our estimates did not consider 
the repulsive force of the nucleus and the decrease of the number of dyneins at the end of the repositioning \cite{hornak_stochastic_2020}, 
see Fig. S4(d-f).  
The MTOC is less pressed against the nucleus when  $\rho_{\textrm{IS}}=80\mu\textrm{m}^{-2}$, see Figs. S4d and e.
\label{fig:estimations}   }
\end{figure}

The MTOC position during the repositioning was traced
in several experiments \cite{kuhn_dynamic_2002,maccari_cytoskeleton_2016,
quann_localized_2009}. 
The MTOC was originally located approximately at
the opposite side of the target cell and its repositioning to the IS
took more than three minutes.
The repositioning was faster in the experiments performed by Yi et al \cite{yi_centrosome_2013} where the MTOC gets to the  IS(distance $<2\mu\textrm{m}$) 
on average in ca 2min.
The increased speed is likely caused by the experimental setup in which the MTOC and the target cell are initially
 diametrically opposed.
In such a configuration all MTs long enough intersect the center of the IS, visually demonstrated in Figs. S1 b and g, and their plus ends attach and are pulled by capture-shrinkage dyneins.
Moreover, all MTs long enough to reach the IS can attach to cortical sliding dyneins.
When $\beta<\pi$, less dyneins are attached to MTs, since just a fraction of MTs intersect the the IS and its center, see Fig. \ref{fig:estimations}c and visually in Fig. S\ref{fig:variable_Beta_basic}.
Consequently, the initial diametrical opposition of the MTOC and the IS may result in higher pulling forces.
\newline

In what follows we estimate the time scales as one would them to be predicted on the basis of our model assumptions.
The drag force acting on a MT moving with velocity $v$ is 
$F_{\rm drag}=\gamma_{\rm MT}\cdot v $, where $\gamma_{\rm MT}$
is the drag coefficient. For a cylindrical object of length $L$
and diameter $d$ it is given by \cite{howard_mechanics_2001}
\begin{equation}
\label{darg_coefficient}
\gamma_{\rm MT}=\frac{4\pi\mu L}{\ln(L/d)+0.84},
\end{equation}
where $\mu$ is the viscosity of the surrounding liquid, the cytoplasm, which 
is $e$ times the viscosity of water, 
$\mu=e\cdot\mu_{w}\approx 10^{-3}{\rm N\,s\,m^{-2}}
=e\cdot 10^{-3}{\rm N\,s\,m^{-2}}$ and we estimate it to be $e\approx 30$ \cite{hornak_stochastic_2020}.
Note that for simplicity we do not discriminate between movement of the cylindrical
object in the longitudinal or in the transverse direction. Taking the average 
length of the MT to be $L=10\mu m$ and its diameter to be $d=25\textrm{nm}$ we
have  $\gamma_{\rm MT}\approx\mu\cdot 18.4\mu m$.
The drag coefficient of the whole cytosceleton with $N_{\rm MT}$ MTs 
the is $\gamma_{\rm cyto}=N_{\rm MT}\cdot\gamma_{\rm MT}$. 
Mitochondria, Golgi apparatus 
\cite{xu_asymmetrical_2013,ladinsky_golgi_1999,day_three-stage_2013,huang_golgi_2017}
and endoplasmic reticulum
\cite{westrate_form_2015,english_endoplasmic_2013,english_peripheral_2009,shibata_rough_2006,
hu_weaving_2011} are massive organelles entangled with the cytoskeleton \cite{gurel_connecting_2014,maccari_cytoskeleton_2016}
and dragged with it, thereby increasing the drag coefficient by a factor 
$g$, i.e. $\gamma_{\rm eff}=g\cdot\gamma_{\rm cyto}$, 
which was estimated to be $g\approx 3$ \cite{hornak_stochastic_2020}.
\newline

The force pulling on the cytoskeleton is given by the number of 
dyneins attached to MTs times the average forces exerted by a dynein
motor: $F=N_{\rm dyn}\cdot F_{\rm dyn}$, the latter is in the pico-Newton 
range, $F_{\rm dyn}=f\cdot 10^{-12}N$, with $f\approx 1$. Consequently, the 
velocity of the whole cytoskeleton movement when $N_{\rm dyn}$ are pulling is
\begin{equation}
\label{speed_estimation}
v=\frac{F_{\rm dyn}}{\gamma_{\rm eff}}
\approx 54\cdot\frac{N_{\rm dyn}}{N_{\rm MT}}\cdot\frac{f}{e\,g}\;
\frac{\mu{\rm m}}{\rm s}.
\end{equation}
Inserting the estimates $f=1$, $e=30$, $g=3$, and evaluating the
r.h.s. for $N_{\rm MT}=100$ MTs and $N_{\rm dyn}=10-50$ attached dyneins 
one obtains a velocity $v= 3.6-18{\mu\rm m}/{\rm min}$, a range that agrees
well with the experimentally determined MTOC velocities \cite{yi_centrosome_2013}. For an initial
MTOC position diametrically opposed to the IS the MTOC would have to travel
a distance $D=\pi R_{\textrm{Cell}}$, where $R_{\textrm{Cell}}$ is the radius of the cell, and with 
$R_{\textrm{Cell}}\approx 5\mu{\rm m}$ and the above velocity estimate, the whole 
relocation process would need 1-4 minutes, which also agrees with 
the experimentally reported relocation times \cite{yi_centrosome_2013}.
\newline

Since the number of attached dyneins is the central quantity determining the 
speed of the relocation process let us relate to the dynein density and the 
attachment rates that we use in our model. For the capture shrinkage 
mechanism we assume dynein to be concentrated in a central region of 
the IS with radius $R_{\rm CIS}=0.4\mu{\rm m}$ (i.e. an area of $0.5\mu{\rm m}^2$ 
and with a dynein density $\rho_{\rm IS}$.
At medium density of $\rho_{\rm IS}=100\mu{\rm m}^{-2}$ we have $50$
dyneins located in this area and since most MTs in our model reach 
this area, they could in principle all be attached: the average 
distance between dyneins is $D_{\rm d2}=\rho_{\rm IS}^{-1/2}=100\textrm{nm}$ 
for the assumed dynein density and the attachment rate is assumed to be
$p_a=5\cdot\exp(-(d_{\textrm{md}}/p_{d}))\textrm{s}^{-1}$ with $d_{\textrm{md}}$ the distance between a MT and a dynein
and $p_{d}=100\textrm{nm}$ one has $p_a\approx 2{\rm s}^{-1}$, implying that attachment
is fast in comparison to the duration of the relocation process.
Actually, in our simulations we observe that initially ca. one quarter of all
MTs get attached to dynein, some of them even attached simultaneously
to two dyneins. Consequently for $\rho_{\rm IS}=100\mu{\rm m}^{-2}$
we have indeed initially 25-50 dyneins attached to MTs, resulting in
an initial MTOC velocity of $v_{\rm MTOC}=9-18{\mu\rm m}/{\rm min}$. 
In the later stage of the relocation process competing forces 
will slow down the MTOC velocity, which will be revealed by the
actual simulations reported below.
\newline

These rough estimates hold for our model framework as well and can be elaborated more on the basis of more detailed model assumptions. First force exerted by attached dynein is assumed to depend on the length of the stalk 
between the attachment and the anchor point  $l_{\textrm{d}}$ and is expressed as 
 $F_{\textrm{d}} = 0$ if  $l_{\textrm{d}} < L_{0}$ and
$ F_{\textrm{d}} = k_{\textrm{d}}( l_{\textrm{d}} - L_{0} )$ otherwise,
where 
$L_{0} = 18\textrm{nm}$ is the length of the relaxed stalk
and $k_{\textrm{d}} = 400\textrm{pN} \mu \textrm{m}^{-1}$
is the elastic modulus of the stalk, see Fig. \ref{fig:estimations}a.
In our model, the dynein makes 
 steps with the length
$d_{\textrm{step}}=8\textrm{nm}$
 towards the minus-end of the MT.
 The stepping is very fast at zero load(the first two steps), it slows down as the force increases \cite{hornak_stochastic_2020} and the movement stops at the stall force  $F_{S}=4$pN.
 Since the MT depolymerizes and moves, the distance between the attachment and the anchor point can differ from the multiples of the  dynein step.
Consequently, the length of the stalk is  $l_{1}<l_{\textrm{d}}<l_{2}$ where  $l_{1}$ and $l_{2}$ are the lengths corresponding to the second step
and to the stall force, respectively, see Fig.
\ref{fig:estimations}a.
The average dynein force $\overline{F}_{\textrm{d}} = 1.66$pN is calculated as the integral of the force between  $l_{1}$ and $l_{2}$ divided by their distance.
\newline

At the beginning of the repositioning, the number of attached dyneins increases fast, see Fig. S2.
The detachment rate of the dynein is 
$ p_{\textrm{det}} = \textrm{exp}(\frac{F_{d}}{F_{D}})s^{-1}$,
where $F_{D} =2$pN.
The 
detachment rate corresponding to the average dynein force is
$p_{\textrm{det}}(\overline{F}_{\textrm{d}}) = 2.29\textrm{s}^{-1}$,
see Fig.
  \ref{fig:estimations}b. 
Consequently, dyneins are expected to detach in less than half a second.
The attachment rate of the dynein decreases exponentially with the distance from the filament \cite{hornak_stochastic_2020}.
When the distance between the dynein and the filament
$d_{\textrm{md}} = 95$nm, the attachment rate equals detachment rate of the 
average dynein force  $p_{a}(d_{\textrm{md}} ) = p_{\textrm{det}}(\overline{F}_{\textrm{d}})$, see Fig. \ref{fig:estimations}b.
Consequently, the dyneins located closer to the filament are expected to attach faster than dyneins detach on average.
The fraction of MTs intersecting the IS, $q_{\textrm{IS}}$, (or the central region of the IS, $q_{\textrm{CIS}}$), shown in Fig. \ref{fig:estimations}c, is given by the ratio of the diameter of the IS (or diameter of the center of the IS) and the circumference $c(\beta)$ of the circle of latitude at angle $\beta$ (see Fig. \ref{fig:variable_Beta_basic}d): 
$q_{\textrm{IS}} = \textrm{min}( 1, 2 R_{\textrm{IS}} / c(\beta))$, with $c(\beta)= 2 \pi r(\beta)$, where $r(\beta) =R_{\textrm{Cell}}\textrm{sin}(\beta)$ is the radius of the circle.

The number of attached dyneins can then  be estimated by the number of dyneins that are closer than $d_{\textrm{md}} = 95\textrm{nm}$ to a MT:
\begin{equation}
\label{number_of_dynein}
\overline{N}_{dm} = N_{\textrm{MT}} \cdot q_{\textrm{CIS}} \cdot n_{dm},
\end{equation}
where $N_{\textrm{MT}} = 100$ is the number of MTs and 
$n_{dm} = \pi \cdot d_{\textrm{md}}^{2} \cdot \rho_{\textrm{IS}}$ is the number 
of dynein in the proximity of the filament. 
It can be seen in Fig. S2 that the number of attached dyneins approaches the estimated number of dyneins regardless of the angle and the dynein density.   
With this $\beta$-dependent estimate of the $\beta$-dependent number of attached dyneins we can perform again the calculation of the estimated MTOC velocity and the relocation time as above.
\newline

Fig. \ref{fig:estimations}d shows 
the dependence of the estimated time of the repositioning
$T_{\textrm{est}}$
 on the
angle $\beta$.
The repositioning time increases with the angle
 until they reach a maximum at $\beta\sim0.6\pi$ and then they decrease,
 see \ref{fig:estimations}d.
The decrease when $\beta\leq0.5\pi$ can be explained by the fact that the 
distance is increasing and the $q_{\textrm{CIS}}$ decreases with $\beta$, see Fig. \ref{fig:estimations}d.
The ratio $q_{\textrm{CIS}}$  
increases sharply when  $\beta>0.65\pi$
and the increased pulling force results in faster repositioning.
This may offer an explanation why the time of
repositioning is the shortest in the experimental setup when the MTOC and the IS are initially diametrically opposed
\cite{yi_centrosome_2013}.

\subsection*{Repositioning with one IS}

\begin{figure}[hbt!]
\centering
     \includegraphics[trim=0 650 0 0,clip,width=0.99\textwidth]{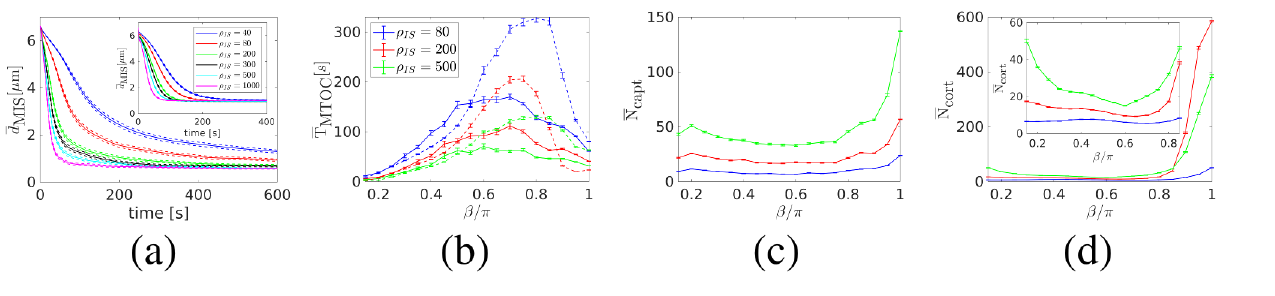}            
\caption{Repositioning under the influence of the capture-shrinkage and cortical sliding mechanisms for different angles $\beta$ between the IS and the initial position of the MTOC.
(a) Dependence of the average MTOC-IS distance $\bar{d}_{\textrm{MIS}}$ on time. $\beta = 0.5\pi$.
(b) Dependence of the averaged times 
$\overline{\textrm{T}}_{\textrm{MTOC}}$ 
(MTOC-IS distance 
$d_{\textrm{MIS}}<1.5\mu\textrm{m}$)
 on the angle $\beta$.
Capture-shrinkage and cortical sliding mechanisms are represented by solid and dashed lines, respectively.
(c-d) Dependencies of the  the mean numbers of attached dyneins $\overline{N}$
averaged over simulation runs  on the angle $\beta$.  
(c) Capture-shrinkage mechanism.
(d) Cortical sliding mechanism.
 \label{fig:variable_Beta}
}
\end{figure}

In the first part of our work \cite{hornak_stochastic_2020}, we analyzed the repositioning in the cell where the MTOC and the IS are initially diametrically opposed.
In this section we present the results of the computer simulation of the model with one IS, located at an angle $\beta$ with respect to the initial position of the MTOC, see Fig. \ref{fig:variable_Beta_basic}.
Fig. \ref{fig:variable_Beta}a shows that expectedly the repositioning becomes faster with increasing dynein density for both mechanisms.
Moreover, the MTOC dynamic has the same characteristic as in the case of 
$\beta = \pi$, which was analyzed in detail in \cite{hornak_stochastic_2020}. 
The MTOC travels to the IS and its speed decreases with the MTOC-IS distance.
Additional analysis of the repositioning for the cases of 
$\beta = 0.75,0.5,0.25\pi$ can be found in Supporting Materials and Methods, Sections 2.1
and 2.2. 
Here we focus on the average repositioning time $\overline{T}_{\textrm{MTOC}}$ and its dependence on the angle $\beta$: Fig. \ref{fig:variable_Beta}b shows
 that
 $\overline{T}_{\textrm{MTOC}}$ increases with the angle $\beta$ to a maximum 
 at $\beta \sim 0.75\pi$
and then decreases. 
 $\overline{T}_{\textrm{MTOC}}$ depends on the initial MTOC-IS distance, opposing forces and the pulling force of dynein motors.
The opposing forces increase with the angle $\beta$ since the nucleus increasingly presents an obstacle on the path of the MTOC.
For $\beta = 0.25\pi$ the nucleus does not intersect the line between the initial positions of the MTOC and the IS, visually demonstrated in
Supporting Materials and Methods, Figs. S1b-k.
Contrarily, the MTOC has to navigate around the entire nucleus when $\beta=\pi$.
\newline

Figs. \ref{fig:variable_Beta}c and d show that the number of attached dyneins decreases with  $\beta$ to a minimum at approximately $\beta = 0.6\pi$ and then increases sharply.
This can easily be explained 
by the number of MTs intersecting the IS given by 
the ratio of the diameter of the IS (or its center) and the circumference of the circle of latitude at angle $\beta$ (see Figs. \ref{fig:variable_Beta_basic}d and \ref{fig:estimations}c): $q_{\textrm{IS}} = \textrm{min}( 1, 2 R_{\textrm{IS}} / c(\beta))$, with $c(\beta)= 2 \pi r(\beta)$, where $r(\beta) =R_{\textrm{Cell}}\textrm{sin}(\beta)$ is the radius of the circle.
In the special case of $\beta=\pi$ all MTs long enough intersect the IS, as visualized in Figs. S1c and h.
However, when $\beta=0.5\pi$ the IS is intersected only by MTs sprouting from the MTOC towards it, visually in Figs.
\ref{fig:variable_Beta_basic}b and d and Figs. S1a and e. 
The ratio decreases with the angle $\beta$ until it reaches the minimum at $\beta=0.5\pi$ and then it increases sharply, see Fig. \ref{fig:estimations}c and visually demonstrated in \ref{fig:variable_Beta_basic}d. 
In the simulations the minimum is slightly shifted from $\beta=\pi/2$ to $0.6\pi$, as visible in Figs. \ref{fig:variable_Beta}c and d, because dyneins detach due to an increasing opposing force of the nucleus.
Subsequently, the number of dynein increases due to the fact that the increasing percentage of MTs intersects the IS,
compare Figs. \ref{fig:variable_Beta}c and d with Fig.\ref{fig:estimations}c.
By comparing Fig. \ref{fig:variable_Beta}c and \ref{fig:variable_Beta}d one observes that 
the number of attached cortical sliding dyneins increases more sharply with increasing $\beta$, due to the fact that a part of the relatively large IS is located in the diametrical opposition of the IS for 
$\beta>0.9\pi$ and the MTs sprouting from the MTOC in all directions can attach to dynein. 
The number of attached capture-shrinkage dyneins at $\beta=0.15\pi$  is smaller than for
$\beta=0.2\pi$  due to the fact that due to the short MTOC-IS distances the MTOC is dragged towards the IS and the number of dyneins quickly decreases, see Fig. S4.
\newline

The repositioning time, $\overline{T}_{\textrm{MTOC}}$, increases with $\beta$ between 0 and $\beta<0.7\pi$, since the distance and opposing force increase and the number of 
attached dyneins decrease, see Fig. \ref{fig:variable_Beta}b-d.
It can be seen in Fig. \ref{fig:variable_Beta}b
that the increase of  $\overline{T}_{\textrm{MTOC}}$ is sharper for cortical sliding when $\beta>0.5\pi$
due to the different behavior of  the number of attached dynein, c.f. 
S4 and S6.
When $\beta>0.8$, $\overline{T}_{\textrm{MTOC}}$ decreases rapidly due to the sharp increase of pulling force.
\newline

The repositioning time offers a way how to compare the performance of the two mechanisms for different configurations of the cell.
It can be seen that the cortical sliding mechanism 
outperforms the capture-shrinkage mechanism when $\beta < 0.5\pi$
and is substantially slower otherwise.
The only exception is the case of cortical sliding mechanism when the density $\tilde{\rho}_{\textrm{IS}} =200\mu\textrm{m}^{-2}$ since  it results in the fastest repositioning when $\beta\geq 0.85\pi$.
The speed of the process can be explained by the three regimes of cortical sliding repositioning analyzed in \cite{hornak_stochastic_2020}.
The difference between the repositioning times for the two mechanisms decreases
as the dynein density increases, see Fig. \ref{fig:variable_Beta}b. 

In our model the cortical sliding dynein was distributed equally over the entire IS. However, we observe that the large majority of attached cortical sliding dynein is located at the periphery of the IS, see Fig. S7 in Supporting Materials and Methods.
 In the case of combined mechanisms, the attached cortical sliding dyneins are completely absent in the center of the IS, see Fig. S9c in Supporting Materials and Methods.
It was hypothesized \cite{kuhn_dynamic_2002} that the dynein colocalizes with the  ADAP ring at the periphery of the IS to fascilitate the interaction with MTs. 
This finding supports the aforementioned hypothesis.

\subsection*{Repositioning in the T Cell with two IS }

In this section we present the results of the computer simulation of the model with two IS, as sketched in Fig. \ref{fig:two_IS_sketch}.
The configuration of the cell is defined by the angle $\gamma$ between the two IS, sketched in Fig.  Fig. \ref{fig:two_IS_sketch}a.
The densities of dyneins anchored at
both IS, $\tilde{\rho}^{1}_{\textrm{IS}}$ and $\tilde{\rho}^{2}_{\textrm{IS}}$, and the central region of the IS,  $\rho^{1}_{\textrm{IS}}$ and $\rho^{2}_{\textrm{IS}}$, are unknown
model parameters, which we therefore vary over a broad
range between 0 (no anchored dynein) and $1000\mu\textrm{m}^{-2}$.
We calculate and analyze the following quantities: the transition frequency between  the two IS, $N_{\textrm{tr}}$ $\textrm{min}^{-1}$; the dwell times
at one IS, which is defined as the time interval during which the
 MTOC-IS distance is smaller than
$3\mu\textrm{m}$, $T_{\textrm{d}}$;  
the longitudinal and transverse fluctuations of the MTOC by determining the time averaged probability distribution of the polar and azimuthal angle, $\varphi$ and $\theta$, respectively, which are defined as sketched in Fig. \ref{fig:two_IS_sketch}b.
For each point in the parameter space, these quantities were averaged over 500 simulation runs. Each simulation run is initialized with  all dyneins being detached. Results are shown
with the standard deviation as error bars.

\subsubsection*{Capture-shrinkage mechanism }
\label{capture_shrinkage_2_IS}

\begin{figure}[hbt!]
\centering
     \includegraphics[trim=0 300 0 0,clip,width=0.666\textwidth]{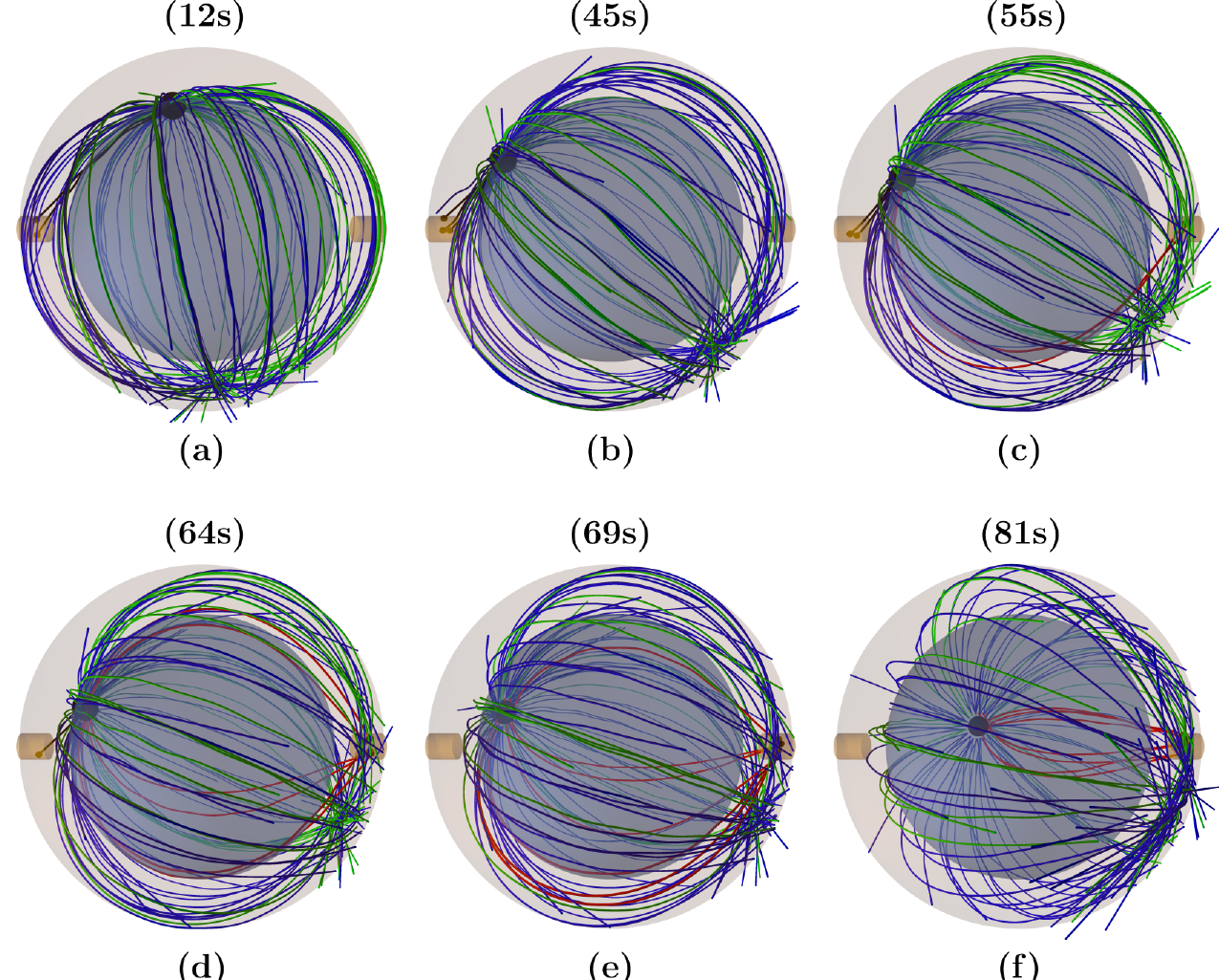}      
   \caption{Repositioning with two IS:  Snapshots from the time evolution of
the MT cytoskeleton configuration with
 capture-shrinkage mechanism,
 $\rho_{IS}^{1} = \rho_{IS}^{2} = 400\mu m^{-2}$. 
The MTOC is indicated by the large black sphere.
Brown cylinders indicate the centers of both IS where capture-shrinkage dyneins are located.
Black and red lines represent MTs attached to capture-shrinkage dyneins
and
blue and green lines indicate growing, shrinking unattached MTs, respectively.
Small black spheres in both IS represent attached dyneins.
(a) MTs attach to dyneins in the left IS, form a stalk and the MTOC moves towards the left IS.
(b) MTOC approaches the IS and MTs depolymerize.
(c) Short MTs detach from the left IS. Simultaneously, the plus end of the MT intersect with the center of the distant IS and are captured by dynein.
(d) Only one MT remains attached to the dynein in the left IS and additional 
MTs attach in the right IS.
(e) All MTs are detached from the left IS and multiple MTs are attached in the distant IS.
(f) MTs stalk is formed and the MTOC moves towards the right IS. 
       \label{fig:two_IS_sketch_capt_capt}} 
 \end{figure}

\begin{figure}[hbt!]
\centering
     \includegraphics[trim=35 410 30 30,clip,width=0.66\textwidth]{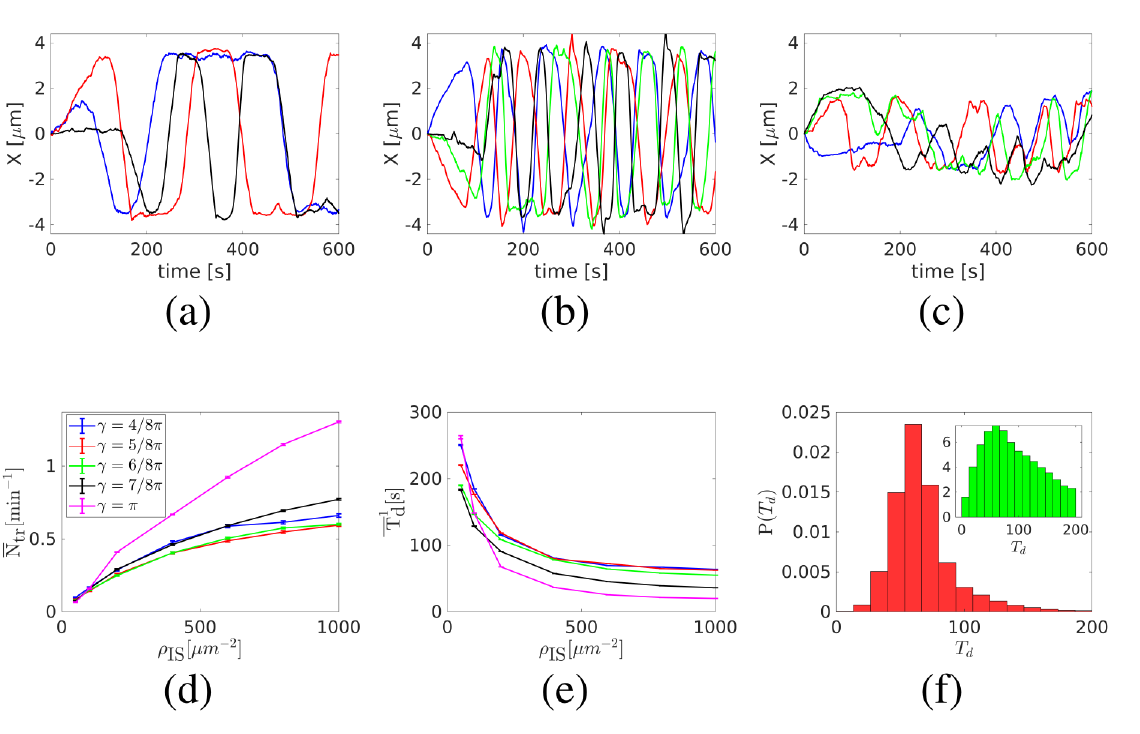}     %
   \caption{Capture-shrinkage mechanism with two IS with the same dynein density $\rho_{IS}^{1} = \rho_{IS}^{2} = \rho_{IS}$.
(a-c) Examples of the time evolution of the MTOC position in 600s of the simulation. The time evolutions of x coordinate of the MTOC are shown. Both IS are located in X-Z plane and the MTOC is originally located at the same distance from both IS, $x = 0$.
(a and b) $\gamma = \pi$. 
(a) $\rho_{IS} = 200\mu\textrm{m}^{-2}$. 
(b) $\rho_{IS} = 800\mu\textrm{m}^{-2}$. 
(c) $\gamma = \frac{1}{2}\pi$, $\rho_{IS} = 800\mu\textrm{m}^{-2}$.
(d) Dependencies of the average transition frequency between two IS per minute
(e), the average dwell time that the MTOC spends next to the IS(f) on dynein density $\rho_{IS}$ are shown. 
(f) Probability densities of dwell times for angle 
between the axis of two IS $\gamma = \pi$, $\rho_{\textrm{IS}} = 200\mu\textrm{m}^{-2}$.
Inset: Dwell time distribution in a log-lin plot demonstrating the exponential tail.
Dwell times were collected from 1280 simulation runs.
       \label{fig:two_IS_capture}} 
 \end{figure}

Movie S1 of the Supporting Materials and Methods shows the repositioning with two IS with the same  density of capture-shrinkage dyneins $\rho_{IS}^{1} = \rho_{IS}^{2} = 400\mu m^{-2}$.
In the first seconds of the simulation, MTS attach to dyneins at the left IS,
visualized in Fig. \ref{fig:two_IS_sketch_capt_capt}a, and the MTOC is dragged towards it.
 Captured MTs 
shorten and depolymerize, see Fig. \ref{fig:two_IS_sketch_capt_capt}b.
As the MTOC approaches the left IS, 
we observe that the number of attached MTs decreases as MTs detach and reattach in the center of the IS, visually demonstrated in Figs. \ref{fig:two_IS_sketch_capt_capt}b-d.
Simultaneously, the plus end of MTs intersect with the distant IS and are captured by dyneins. 
Finally, all MTs are detached from the left IS at the end of the transition, 
and the MTOC moves to the right IS, visually demonstrated in Figs. \ref{fig:two_IS_sketch_capt_capt}e and f.
Due to the dynamic instability, MTs grow(blue lines) and shrink(green lines).
The MT cytoskeleton is not damaged permanently by the capture-shrinkage mechanism since
short filaments regrow due to the polymerization.
Therefore, the MTOC relocates back and forth between the two IS until one IS is removed. 
In Figs. \ref{fig:two_IS_capture}a-c we see
that the time evolution of the MTOC position
follows a recurring pattern already seen in the movie: 
the MTOC travels to one IS, remains in its close proximity for a time
and then repositions to the second IS.
This pattern is caused by two effects: the MTs lose contact with the close IS(to which it moves)
and establish a contact with the distant IS.
The similar process  for the case $\gamma=\frac{3\pi}{4}$ is shown in Movie S2 of the Supporting Materials and Methods.
\newline

The physical mechanism underlying the MTOC transition from one IS to the other and back is the decrease of the dynein attachment probability with decreasing MTOC-IS distance due to strong bending of attached filaments at short distances as sketched in Fig. \ref{fig:two_IS_sketch}d:
The MTOC is a planar structure and MTs sprout from the MTOC tangentially.
At large MTOC-IS distances the MT bends around the nucleus
and  bending forces press the plus-end against the cell membrane
where it can be captured by dyneins.
At small MTOC-IS distances an attached MT has to bend to stay in
a contact with the IS. 
When a short MT detaches from dyneins, the plus end recedes from the 
IS making a reattachment unlikely.
The attachment probability of the MT in the IS depends on the circumferential MTOC-IS distance,
since only MTs having a length roughly corresponding to it can attach in the IS. 
Fig. \ref{fig:two_IS_sketch}(e) shows that the probability density of the MT length steadily increases before reaching a peak at $L_{\textrm{MT}} \sim 15.8\mu\textrm{m}$ corresponding to the circumferential distance 
between two IS when $\gamma = \pi$.
Consequently, the probability of the MT attachment in the distant IS increases as the MTOC recedes
since increasing number of MT have a length corresponding to the circumferential MTOC-IS distance. 

By comparing Figs. \ref{fig:two_IS_capture}a and \ref{fig:two_IS_capture}b one realizes 
that when the density increases, 
transitions are faster and 
the MTOC remains close to the IS for a shorter time.
Fig. \ref{fig:two_IS_capture}d shows that the  transition frequency increases with the dynein density.
Dwell times decrease  with the increasing dynein density and increasing angle $\gamma$,
c.f. Fig. \ref{fig:two_IS_capture}e. 
One would expect that the transition frequency decreases with the rising distance
between two IS(increasing with $\gamma$).
Surprisingly, it decreases with $\gamma$ only when  $\gamma\leq\frac{3\pi}{4}$ and is the maximal when $\gamma=\pi$.
\newline

The dynein detachment probability is force dependent and its pulling force is constantly opposed by forces of the friction 
 and from the nucleus.
As the density increases, more dyneins share the load from opposing forces and the detachment 
probability decreases leading to shorter dwell times, c.f. Fig. \ref{fig:two_IS_capture}e.
Fig. \ref{fig:two_IS_capture}f shows the probability distribution of dwell times and from the log-lin scale of the same plot in the inset one concludes that the dwell time distribution has an exponential tail. 
An increased dynein number leads to faster MTOC movement and shorter dwell times, which again result in an increased transition frequency.
\newline

The transition frequency does not decrease monotonously with increasing angle $\gamma$ since the probability of dynein attachment increases with the circumferential distance between two IS.
At the end of the MTOC transition, 
only MTs having a length roughly corresponding to the circumferential distance between two IS  can attach at the distant IS, as visualized in Fig. \ref{fig:two_IS_sketch_capt_capt}.
The MT length distribution increases until it reaches maximum corresponding approximately to the half of the cell circumference,
c.f. Fig. \ref{fig:two_IS_sketch}f.
Consequently, the probability that a plus end intersects with the center of the IS at the end of MTOC transitions increases with the angle 
$\gamma$.
The transition frequency decreases with the angle when $\gamma\leq\frac{3\pi}{4}$ because the MTOC travels 
longer distances and the increase of probability density is not significant.
When $\gamma>\frac{3\pi}{4}$ the increasing distance is compensated by  a higher number of MTs intersecting with the center of the distant IS
 leading to shorter dwell times and faster MTOC movement,
 see Figs. \ref{fig:two_IS_capture}b-e.
The case of $\gamma = \pi$ has the additional geometrical advantage that 
all MTs with sufficient length intersect the distant IS at the end of  transitions, visually demonstrated in
Fig. \ref{fig:two_IS_sketch_capt_capt}(b).
\newline

The increasing number of attached MTs influences the continuity of the MTOC transitions.
When $\gamma = \pi$, the movement of the MTOC is regular and uninterrupted, see
Fig.  \ref{fig:two_IS_capture}b.
On the other hand, for the smallest value of $\gamma$, i.e. the shortest distance between the two IS,  the movement of the MTOC is highly irregular, see Fig. \ref{fig:two_IS_capture}c: the MTOC stops and stalls before resuming the movement to the IS (blue, green). 
In some cases the MTOC does not finish the journey and returns to the original location (black).
When $\gamma = \pi$ a relatively high number of MTs intersects with the center of
the distant IS with their plus end, see Fig. \ref{fig:two_IS_sketch}f.
Since the MTOC is pulled by dyneins acting on multiple MTs, transitions are smooth and uninterrupted. 
When $\gamma = \pi/2$ only a limited number of MTs is pulled resulting in  easily interrupted transitions.
\newline

\begin{figure}[hbt!]
\centering
     \includegraphics[trim=35 410 30 30,clip,width=0.66\textwidth]{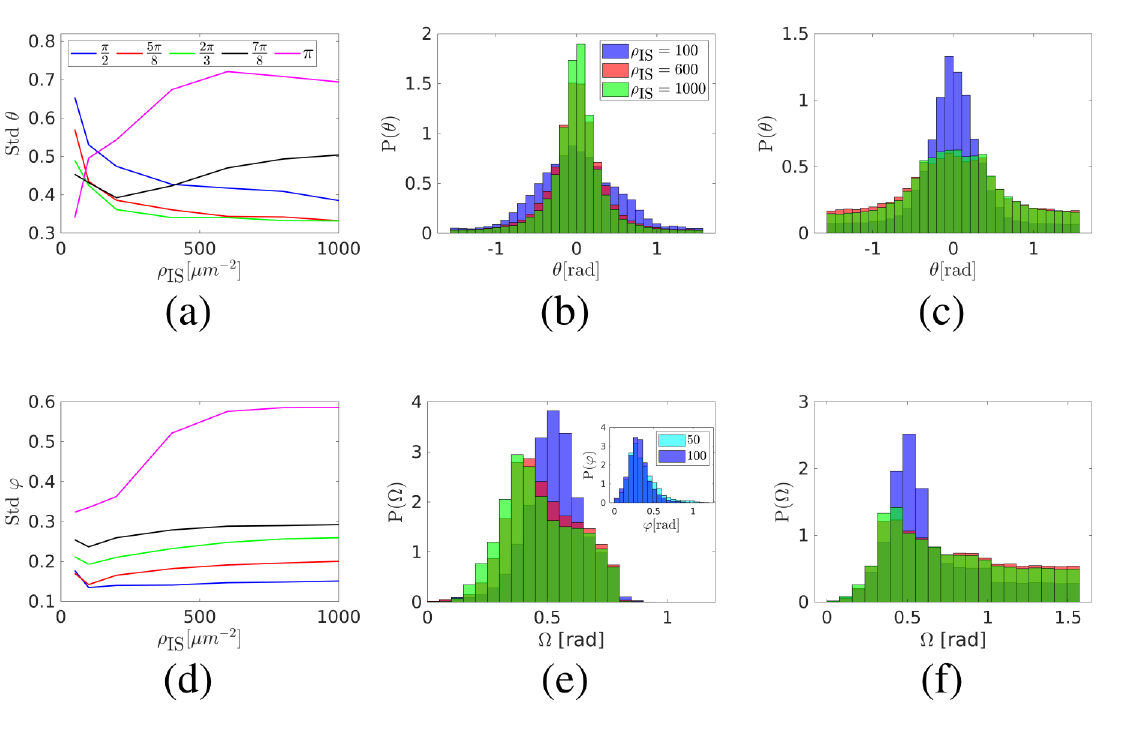}        
   \caption{Capture-shrinkage mechanism in the cell  with two IS with the same dynein density$\rho_{IS}^{1} = \rho_{IS}^{2} = \rho_{IS}$.
(a) The dependence of the standard deviation of the azimuthal angle $\theta$  
on the dynein density  $\rho_{IS}$.
(b and c) Probability densities of the azimuthal angle $\theta$.  
(b) $\gamma = \frac{1}{2}\pi$.
(c) $\gamma = \pi$.
(d)  The dependence of the standard deviation of the polar angle $\varphi$ on the dynein density $\rho_{IS}$.
(e and f) Probability densities of the MTOC-IS angle $\Omega$.  
(e) $\gamma = \frac{1}{2}\pi$. The probability density of the polar angle 
$\varphi$ is shown in the inset.  
(f) $\gamma = \pi$.
       \label{fig:two_IS_capture_2_angle}} 
 \end{figure}

The longitudinal and transverse fluctuations of the MTOC along its path from one IS to the other can be described by the distribution of the polar and azimuthal angle,
$\varphi$ and $\theta$,
sketched in Fig. \ref{fig:two_IS_sketch}b.
The standard deviation of the azimuthal angle decreases with the increasing dynein density when $\gamma \leq \frac{3 \pi}{4}$
and increases when   $\gamma = \pi$, see Fig. \ref{fig:two_IS_capture_2_angle}a.
Two forces act on cytoskeleton: forces of dynein pulling the MTOC towards the IS and the  forces of the tips
of growing MTs on the cell membrane pushing the MTOC to all directions,
sketched in Fig. \ref{fig:two_IS_sketch}e.
When  $\gamma < \frac{3 \pi}{4}$, only a small fraction of MTs 
sprouting from the MTOC intersect the distant IS at the end of the transition, see Fig.  \ref{fig:two_IS_sketch}f.
Since the stalk pulls the MTOC either within the $xz$ plane or towards it, sketched
in Fig. \ref{fig:two_IS_sketch}e, the azimuthal angle can only decrease during transitions.
At the end of the transition, the dynein detach and forces from growing MT tips can push 
the MTOC from the $xz$ plane,
sketched in Fig. \ref{fig:two_IS_sketch}e, increasing the azimuthal angle. 
Consequently, the standard deviation of the azimuthal angle decreases with  dwell times and therefore decreases with dynein density, see Figs. \ref{fig:two_IS_capture}e and \ref{fig:two_IS_capture_2_angle}a.
Fig. \ref{fig:two_IS_capture_2_angle}b shows  that when $\gamma = \frac{\pi}{2}$ the peak of the probability distribution of the azimuthal angle is located at $\theta = 0$ and narrows for higher dynein densities resulting in a reduced standard deviation. 
When $\gamma \geq \frac{7}{8}\pi$ the transitions can increase the azimuthal 
angle of the MTOC since the MTs sprouting in multiple directions can attach to the IS, as visualized in Fig. \ref{fig:two_IS_sketch_capt_capt}. 
In contrast to the case $\gamma < \frac{7}{8}\pi$, the azimuthal angle increases as the dwell time decreases
when $\gamma = \pi$ since the azimuthal angles are low when the MTOC is in the proximity of
the IS and the transitions pulls it from the plane increasing azimuthal angles of the MTOC, see Fig. \ref{fig:two_IS_capture_2_angle}a and c.
\newline

When  $\gamma <  \pi$ the standard deviation of the polar angle slightly decreases with the dynein density  when $\rho_{\textrm{IS}}<100\mu\textrm{m}^{2}$ and then it increases, see Fig. \ref{fig:two_IS_capture_2_angle}d.
The standard deviation of the polar angle depends on its range.
When $\rho_{\textrm{IS}}\geq 100\mu\textrm{m}^{2}$ the MTOC transitions between two IS, see Fig. \ref{fig:two_IS_capture}d and e and the rising dynein force pulls the MTOC closer to the IS, see Fig.\ref{fig:two_IS_capture_2_angle}e,  increasing the range of the polar angle.
The density of $\rho_{\textrm{IS}}= 50\mu\textrm{m}^{2}$ is an exceptional case since the MTOC does not transition, see 
Fig. \ref{fig:two_IS_capture}d, since the dynein density is too small
to establish the MT stalk.
Consequently, forces from the growing MTs can push the MTOC from both IS increasing the polar angle, see the inset of Fig.  \ref{fig:two_IS_capture_2_angle}e.
Obviously, the standard deviation of the polar angle increases with $\gamma$, see Fig. \ref{fig:two_IS_capture_2_angle}d.
When $\gamma=\pi$, the standard deviation is the largest and increases monotonously with dynein density. 
When $\gamma<\pi$ dyneins always pull the MTOC to the IS located in upper hemisphere, sketched in Fig. \ref{fig:two_IS_sketch}b.
When $\gamma=\pi$ the MTOC can travel
through the lower hemisphere thus increasing the range of the polar angle.
Consequently, the standard deviation of the polar angle increases with the transition frequency, compare Figs. \ref{fig:two_IS_capture_2_angle}d
and \ref{fig:two_IS_capture}d.
Similarly to the case $\gamma<\beta$ the MTOC is pulled closer to the IS
as the dynein density increases,
see Fig. \ref{fig:two_IS_capture_2_angle}f.
\newline

\begin{figure}[hbt!]
\centering
     \includegraphics[trim=35 600 30 30,clip,width=0.666\textwidth]{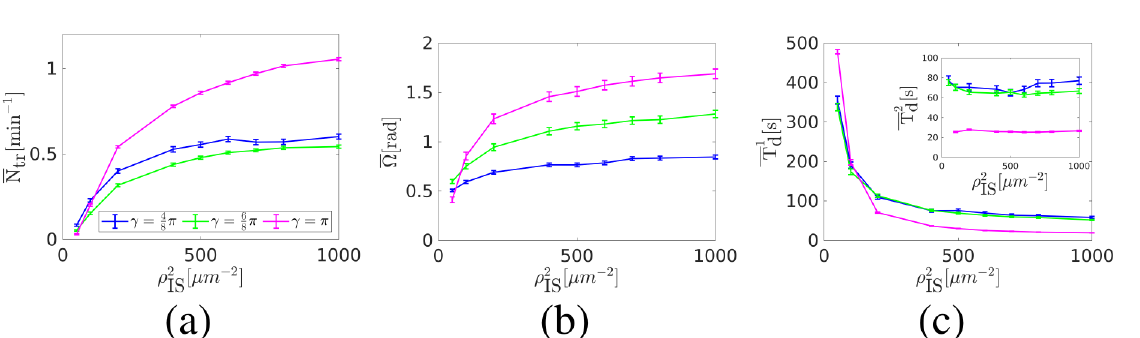}        
   \caption{Capture-shrinkage mechanism in the cell with two IS with different dynein densities.
The dynein density in one IS is fixed, $\rho_{IS}^{1} = 600\mu\textrm{m}^{-2}$ and the density in the second IS ranges between 
$50\mu\textrm{m}^{-2}\leq\rho_{IS}^{2}\leq 1000\mu\textrm{m}^{-2}$.  
Results were collected from 280 simulation runs.
(a-c) Dependencies of the average transition frequency per minute
$\overline{N}_{\textrm{tr}}$ (a), 
the average MTOC-IS angle  $\overline{\Omega}$ (b),
and the average dwell times close to the first IS, $\overline{T}_{d}^{1}$,
and the second IS  $\overline{T}_{d}^{2}$(c)
on the dynein density in the second IS $\rho_{IS}^{2}$ are shown.
       \label{fig:two_IS_capture_2_unequal}} 
 \end{figure}

We further analyzed the capture-shrinkage scenario with different dynein densities at the two IS.
As in the case of equal densities,
dynein detaches when the MTOC approaches the IS and MTs attach at the second IS, visually in Fig. \ref{fig:two_IS_sketch_capt_capt}b.
We fixed the density in $\textrm{IS}_{1}$ $\rho_{\textrm{IS}}^{1}=600\mu\textrm{m}^{-2}$ and we vary the density in $\textrm{IS}_{2}$ $50\mu\textrm{m}^{-2}<=\rho_{\textrm{IS}}^{2}<=1000\mu\textrm{m}^{-2}$. 
From Fig. \ref{fig:two_IS_capture_2_unequal} one can see that the MTOC transitions between two IS even when dynein densities are different.
The MTOC is predominantly located closer
to the IS with higher dynein density. 
Average MTOC-$\textrm{IS}_{1}$ angle $\overline{\Omega}$
steadily increases with $\rho_{\textrm{IS}}^{2}$
and 
$\overline{\Omega}=\frac{\gamma}{2}$ when $\rho_{\textrm{IS}}^{2} = \rho_{\textrm{IS}}^{1}$, see Fig. \ref{fig:two_IS_capture_2_unequal}b.
Moreover, the dwell times are substantially larger for the IS with
higher density, see Fig. \ref{fig:two_IS_capture_2_unequal}c.
\newline

\subsubsection*{Cortical sliding mechanism }

\begin{figure}[hbt!]
\centering
     \includegraphics[trim=0 540 0 0,clip,width=0.66\textwidth]{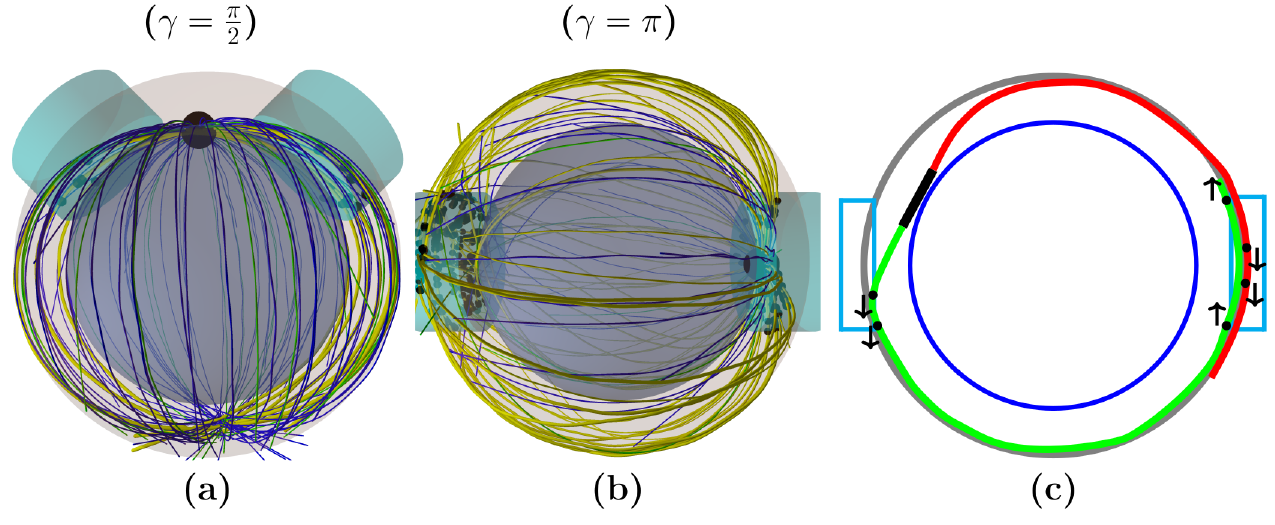}        
   \caption{(a-b) Snapshots from the time evolution of
the MT cytoskeleton configuration with the
 cortical-sliding mechanism acting at both IS. 
MTs are connected to
the MTOC indicated by the large black sphere.
Yellow lines represent MTs attached to cortical-sliding dyneins
and
blue and green lines indicate growing and shrinking unattached MTs, respectively.
Attached cortical-sliding dyneins indicated by small  black spheres  are located in both IS represented by cyan cylinders.
 (a) $\gamma = 0.5\pi$,  $\tilde{\rho}_{IS}^{1} = \tilde{\rho}_{IS}^{2}=\tilde{\rho}_{IS}=200\mu m^{-2}$. MTs attach in both IS and dyneins remain in a thug-of-war for the rest of the simulation. 
(b) $\gamma = \pi$,  $\tilde{\rho}_{IS}^{1} = \tilde{\rho}_{IS}^{2}=\tilde{\rho}_{IS}=1000\mu m^{-2}$. The MTOC is located close to the center of the IS. Almost all MTs are attached to dyneins and they sprout from the MTOC in all directions. 
(c) The sketch of dynein forces when $\gamma = \pi$ and with the MTOC in the close proximity of the IS is shown.
The cell membrane and the nucleus are represented by the gray and blue circle,
respectively.
The black line and the cyan rectangles denote the MTOC and two IS, respectively.
Black dots and arrows denote dynein motors and directions of pulling forces, respectively. 
The green MT is attached to dyneins in both IS and the red MT is attached only by dyneins in the distant IS.
Dyneins in the distant IS pull MTs in different directions.
The MTOC stays in the close proximity of the IS since it is
pulled there by the combined forces of both IS.
       \label{fig:two_IS_sketch_cort_cort}} 
 \end{figure}

In contrast  to the capture-shrinkage mechanism, cortical sliding dyneins are distributed in a relatively large IS and can attach at any position on a MT.
  Since multiple filaments intersect with the IS in every instant, MTs are always simultaneously attached at both IS, as visualized in Figs. \ref{fig:two_IS_sketch_cort_cort}a and b.
\newline

\begin{figure}[hbt!]
\centering
     \includegraphics[trim=35 100 30 30,clip,width=0.66\textwidth]{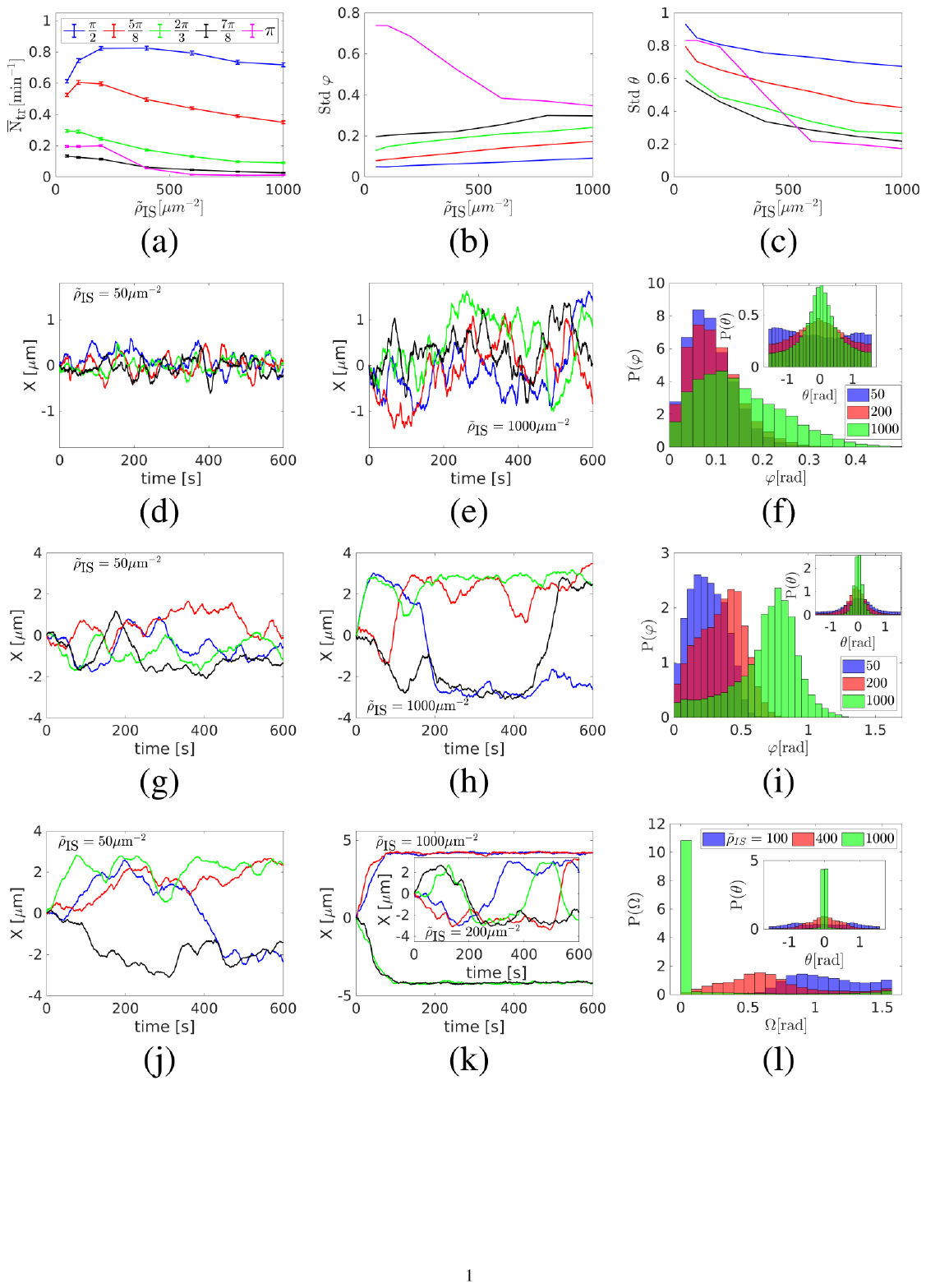}      
   \caption{Cortical-sliding mechanism with two IS with the same dynein density $\tilde{\rho}_{\textrm{IS}}^{1} = \tilde{\rho}_{\textrm{IS}}^{2}=\tilde{\rho}_{\textrm{IS}}$.
(a-c) Dependencies of  
the average transition frequency $\overline{N}_{\textrm{tr}}$  (a),
the standard deviation of the polar angle $\varphi$ (b), and
 the standard deviation of the azimuthal angle $\theta$ (c) on dynein density $\tilde{\rho}_{IS}$ are shown.
(d,e): Examples of the time evolution of the MTOC position for $\gamma=\pi/2$. The time evolutions of x coordinate of the MTOC are shown, (d) for 
$\tilde{\rho}_{\textrm{IS}}=50\mu \textrm{m}^{-2}$, (e) for $\tilde{\rho}_{\textrm{IS}}=1000\mu \textrm{m}^{-2}$. (f) Probability distribution of the polar angle $\varphi$ (main plot) and the azimuthal angle $\theta$ (inset). 
(g-i) The same as (d-e) for  $\gamma=\frac{2 \pi}{3}$. 
(j,k): Examples of the time evolution of the MTOC position for  $\gamma=\pi$, (j) for 
$\tilde{\rho}_{\textrm{IS}}=50\mu \textrm{m}^{-2}$, 
(k) for $\tilde{\rho}_{\textrm{IS}}=1000\mu \textrm{m}^{-2}$ 
(main plot) and 
$\tilde{\rho}_{\textrm{IS}}=200\mu \textrm{m}^{-2}$ (inset). 
(l) Probability distribution of the MTOC-IS angle $\Omega$ and the azimuthal angle (inset). 
  \label{fig:two_IS_cortical}} 
 \end{figure}

By comparison of Figs. \ref{fig:two_IS_cortical}d-l one realizes 
that as the angle $\gamma$  increases,
the MTOC transitions become more continuous and less frequent. 
When $\gamma<\frac{3\pi}{4}$,
the transition frequency increases before reaching the peak at  $\tilde{\rho}_{\textrm{IS}}=200\mu\textrm{m}^{-2}$ and then it declines,  see Fig. \ref{fig:two_IS_cortical}a. 
It steadily decreases with the rising dynein density when $\gamma>\frac{3\pi}{4}$. 
The case of $\gamma=\pi$ is unique since the transition frequency decreases to zero.
Moreover, it is the only case when standard deviation of the polar angle decreases with rising dynein density,  see Fig. \ref{fig:two_IS_cortical}b.
The standard deviations of the azimuthal angle steadily decreases
with the dynein density, see Fig. \ref{fig:two_IS_cortical}c.
\newline

When $\gamma <\frac{3\pi}{4}$ relatively large IS are located close to each other, visually demonstrated in Fig. 
\ref{fig:two_IS_sketch_cort_cort}a.
The dynein detach when the MTOC approaches the IS,
see Section 2.2 of the Supporting Materials and Methods, and the pulling force decreases. MTs are being pulled by dyneins in the second IS at the same time. These two effects result in minimal MTOC fluctuations around the central position, see Fig. \ref{fig:two_IS_cortical}d, and in a relatively high transition frequency between two hemispheres, see Fig. \ref{fig:two_IS_cortical}a.
By comparison of azimuthal angles in Figs. \ref{fig:two_IS_cortical}f and i one realizes that the MTOC fluctuations have a strong lateral component when $\gamma \leq\frac{3\pi}{4}$ which is stronger than the parallel one when $\tilde{\rho} = 50\mu\textrm{m}^{-2}$.
This is due to the fact that dyneins located at the peripheries 
of both IS  can cooperate while pulling the MTOC from the  $xz$ plane but are always in competition when pulling the MTOC parallel to the plane.
The MTOC movement gets more aligned with the $xz$ plane,
see Fig. \ref{fig:two_IS_cortical}f, 
as the dynein density increases leading to a slight increase in the
transition frequency, see Fig. \ref{fig:two_IS_cortical}a. 
As the density further increases, $\tilde{\rho} > 200\mu\textrm{m}^{-2}$, the MTOC is increasingly pulled from the central position to the IS, see Fig. \ref{fig:two_IS_cortical}e and f.
The number of transition decreases since the MTOC travels longer distance.
Moreover, as the MTOC approaches one IS, the forces of nucleus oppose the movement to the distant IS giving the advantage to the dynein at the close IS in the constant tug of war.
Since the nucleus increasingly presents and obstacle between two IS as $\gamma$ increases,
the transition frequency decreases more significantly with densities
 $\tilde{\rho} > 200\mu\textrm{m}^{-2}$ when $\gamma=\frac{5\pi}{6}$,
see Fig.  \ref{fig:two_IS_cortical}a.
\newline

When $ \frac{3\pi}{4}\leq \gamma < \pi$ and dynein densities are low, the constant competition between dyneins from both IS leads to short, interrupted transitions between two IS, see 
Fig. \ref{fig:two_IS_cortical}g.
The MTOC moves around the central position(green), transitions between two IS are very slow(blue), interrupted(red) or the MTOC dwells in one hemisphere for a long time(black).
As in the previous case, the MTOC is increasingly pulled from the central position to the IS with rising density, see Fig. \ref{fig:two_IS_cortical}h and i.
Transitions to the distant IS gets more unlikely due to the fact that the dyneins from the distant IS are opposed by the forces of dyneins from the close IS and from the nucleus.
When $\tilde{\rho} = 1000\mu\textrm{m}^{-2}$ the MTOC dwells in one hemisphere and rarely transitions,
see Fig. \ref{fig:two_IS_cortical}h and i.
Since the MTOC stays longer in the proximity of the IS located at the $xz$ plane as the density increases, 
the peak of azimuthal angle probability distributions narrows, see 
Fig. \ref{fig:two_IS_cortical}i.
The Movie S3  of the Supporting Materials and Methods
shows the process for the case $\gamma=\frac{3}{4}\pi$ and 
$\tilde{\rho} = 600\mu\textrm{m}^{-2}$.
\newline

In Figs. \ref{fig:two_IS_cortical}j and k it can be seen that  the MTOC trajectories are fundamentally different for lower and higher densities when
$\gamma = \pi$.
Moreover, the transition frequency is higher than in the case of 
$\gamma = \frac{7\pi}{8}$, since the higher number of MTs intersect the distant IS when IS are in diametrical opposition,
visually demonstrated in Fig. \ref{fig:two_IS_sketch_cort_cort}.
 When the density is low, the MTOC transitions between two IS never reaching their center, see Fig. \ref{fig:two_IS_cortical}j and l.
As the density increases, the MTOC approaches  the IS closer, see Figs. \ref{fig:two_IS_cortical}k and l. 
When $\tilde{\rho}_{\textrm{IS}}>600\mu\textrm{m}^{-2}$, dynein forces are strong enough to pull the MTOC to the center of the IS,
where it remains for the rest of the simulation, see Fig. \ref{fig:two_IS_cortical}k and l.
In such a case the majority of MTs are attached in the distant IS, visually demonstrated in Fig. \ref{fig:two_IS_sketch_cort_cort}b. 
Since MTs attached at the distant IS are sprouting from the MTOC in every direction, the dyneins act in a competition, sketched in Fig. \ref{fig:two_IS_sketch_cort_cort}c. 
If the MTOC recedes from the center of the IS, the dyneins at the close IS pulls the MTOC back alongside the part of the dynein in the distant IS. 
Contrarily to the cases when $\beta<\pi$ the MTOC can travel to the distant IS in all directions resulting in substantial deviations from the $xz$ plane, see the inset of Fig. \ref{fig:two_IS_sketch_cort_cort}l.
The peak of the probability density of the angle $\theta$ gets more narrow with rising density since the MTOC is increasingly located closer to the IS, \ref{fig:two_IS_sketch_cort_cort}l.
The Movie S4  of the Supporting Materials and Methods
shows the process for the case $\gamma=\pi$ and 
$\tilde{\rho} = 1000\mu\textrm{m}^{-2}$.
\newline

In general, the MTOC is located closer to the $xz$ plane as the density increases, see Figs. \ref{fig:two_IS_cortical}f, i and l.
Consequently,  the standard deviation of the horizontal angle decreases with the dynein density, see Fig. \ref{fig:two_IS_cortical}c.
At the same time the standard deviation of the polar angle increases, see  Fig. \ref{fig:two_IS_cortical}b, due to its increased range, see Figs. \ref{fig:two_IS_cortical}f and i. 
The only exception is the case of $\gamma=\pi$. Small and still decreasing transition frequency cause decreasing range 
MTOC-$\textrm{IS}_{1}$ angles,
see Fig. \ref{fig:two_IS_cortical}a and l, leading to the decreased standard deviation of the polar angle, see 
Figs. \ref{fig:two_IS_cortical}b.

\subsubsection*{Capture-shrinkage and cortical sliding mechanisms in different IS}

\begin{figure}[hbt!]
\centering
     \includegraphics[trim=0 290 0 0,clip,width=0.66\textwidth]{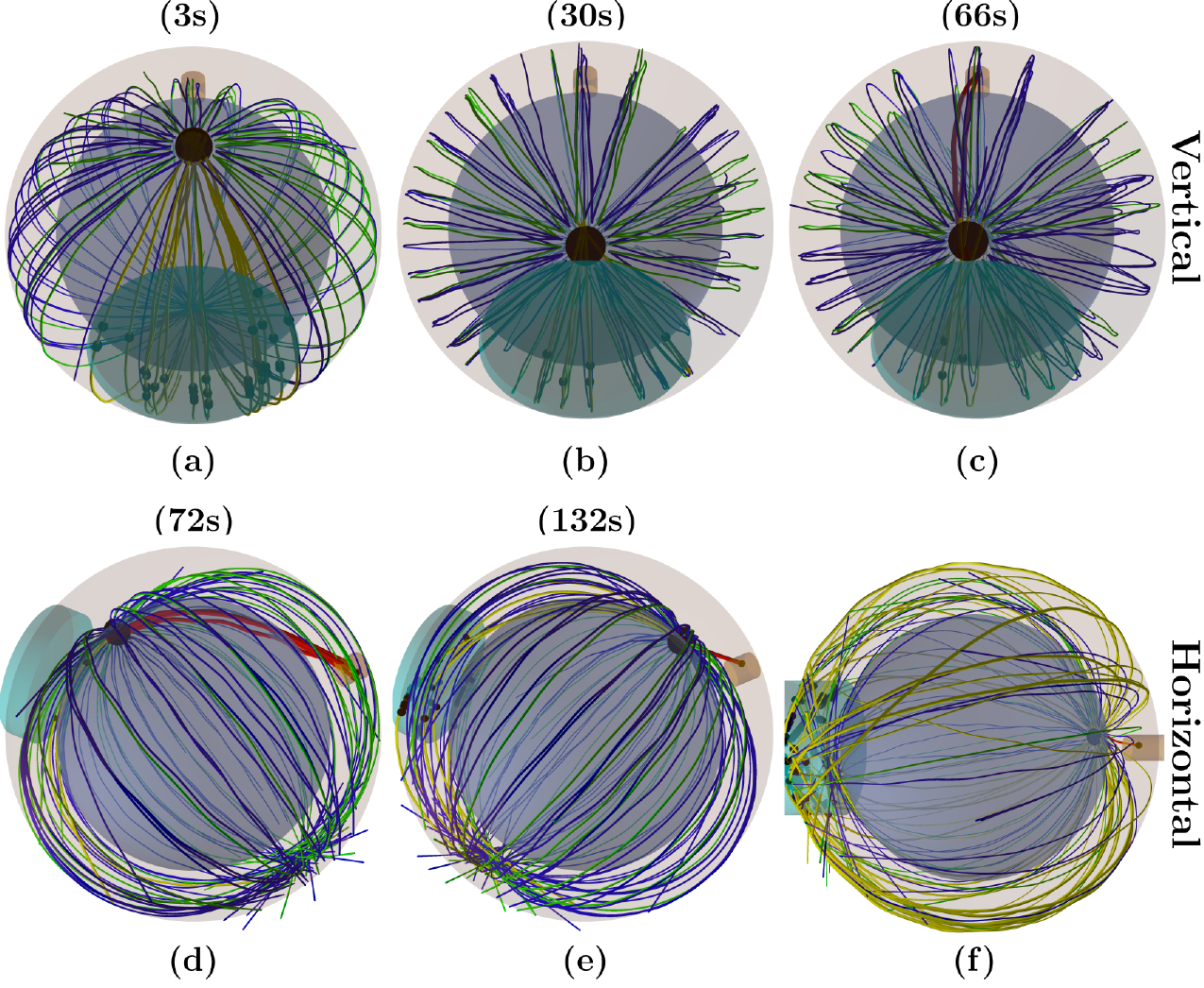}        
   \caption{Snapshots from the time-evolution of the MT cytoskeleton configuration under the effects of both  capture-shrinkage and cortical sliding mechanisms from two perspectives in different IS with the same dynein densities,
$\gamma = \frac{3\pi}{4}$, $\rho_{\textrm{IS}}^{1} = \tilde{\rho}_{\textrm{IS}}^{2} = \rho = 400 \mu\textrm{m}^{-2}$,
$\rho_{\textrm{IS}}^{2} = \tilde{\rho}_{\textrm{IS}}^{1} = 0 \mu\textrm{m}^{-2}$.
 The brown cylinder indicates the center of the  IS where capture-shrinkage dyneins are located and the cyan cylinder indicates the whole IS containing cortical sliding dyneins. Attached dyneins are represented by small, black spheres.
Red and yellow lines represent MTs attached to capture-shrinkage and cortical sliding dyneins, respectively. Blue and green lines depict growing and shrinking unattached MTs, respectively, connected to the MTOC represented by the large, black sphere.
(a) Initially, MTs attach only to the cortical sliding dynein at the left IS since no plus end of MTs intersect with the center of the right IS.  
(b) Cortical sliding dyneins detach as the MTOC approaches the left IS.
(c) The plus end of a MT intersects with the center of the right IS and is captured by dynein.
(d)
Several MTs are still attached to cortical sliding dyneins at the left IS
and
 multiple MTs form a stalk connecting the center of the right IS with the 
MTOC.
 Pulling force of capture-shrinkage dyneins overpowers cortical sliding dyneins and the MTOC moves to the right IS. 
(e) As the MTOC approaches the right IS, capture-shrinkage MTs detach from the dyneins. Simultaneously, MTs attach to cortical sliding dyneins at the left IS resulting in the transition to the left IS.
(f)
Snapshots from the time-evolution of the MT cytoskeleton,
 $\rho = 1000\mu\textrm{m}^{-2}$, $\gamma = \pi$.
The MTOC is located close to the center of the right IS. Almost all MTs are attached to cortical sliding dyneins at the left IS and they sprout from the MTOC in all directions.
The MTOC dwells close to the right IS since the forces from cortical sliding dyneins are pulling the MTOC to different directions and the MTs attached to capture-shrinkage dynein pull the MTOC towards the right IS.  
       \label{fig:sketch_two_IS_combined_one_each_side}} 
 \end{figure}
\begin{figure}[hbt!]
\centering
     \includegraphics[trim=0 250 10 0,clip,width=0.66\textwidth]{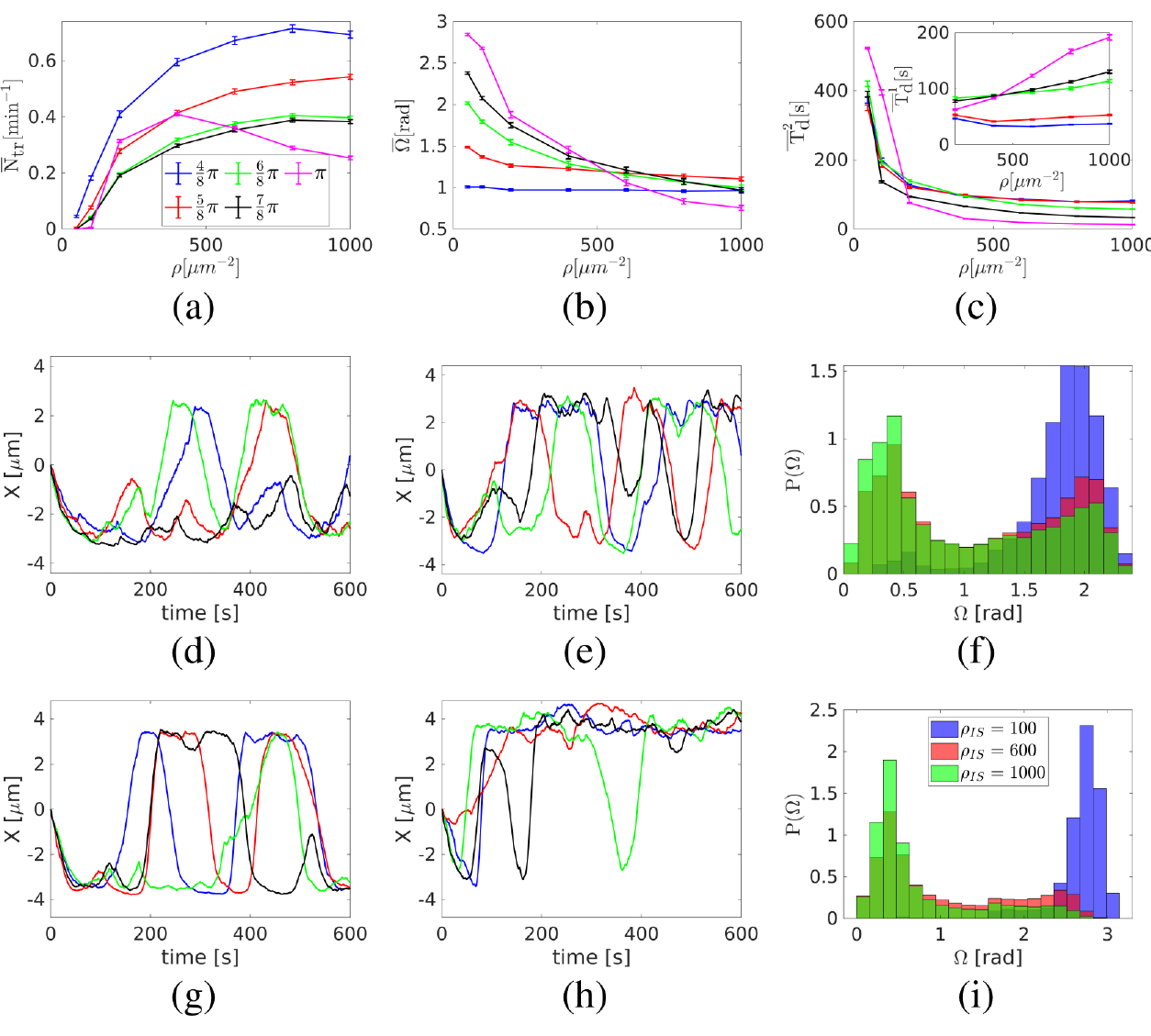}     
   \caption{Capture-shrinkage and cortical sliding mechanisms  in different IS with the same dynein densities $\rho_{\textrm{IS}}^{1} = \tilde{\rho}_{\textrm{IS}}^{2} = \rho$,
$\rho_{\textrm{IS}}^{2} = \tilde{\rho}_{\textrm{IS}}^{1} = 0 \mu\textrm{m}^{-2}$. Cortical sliding $\textrm{IS}_{2}$ is located in the hemisphere $x<0$.
(a-c) Dependencies of  
the average transition frequency $\overline{N}_{\textrm{tr}}$ between two IS (a),
the average angle between the MTOC and the capture-shrinkage $\textrm{IS}_{1}$ $\overline{\Omega}$ (b), and
 average dwell times $\overline{\textrm{T}}^{1}_{\textrm{d}}$ and $\overline{T}^{2}_{\textrm{d}}$ 
 that the MTOC spends next to the capture-shrinkage $\textrm{IS}_{1}$
 and the cortical sliding $\textrm{IS}_{2}$, respectively (c),
 on the dynein density $\rho$ are shown.
The  dwell times $\overline{\textrm{T}}^{1}_{\textrm{d}}$
are given for $\rho \geq 200 \mu\textrm{m}^{-2}$
since the MTOC does not reach the $\textrm{IS}_{1}$ when
 $\rho < 200 \mu\textrm{m}^{-2}$.
(d-e)
Examples of the time evolution of the MTOC position in 600s of simulation.
The time evolutions of x coordinate of the MTOC are shown, $\gamma = \frac{3 \pi}{4}$.
(d) $\rho = 200\mu\textrm{m}^{-2}$.
(e) $\rho = 1000\mu\textrm{m}^{-2}$.
(f) Probability distribution of the angle $\Omega$ between the MTOC and the 
capture-shrinkage $\textrm{IS}_{1}$, $\gamma = \frac{3\pi}{4}$.
(g-h)
Examples of the time evolution of the MTOC position in 600s of simulation are shown, $\gamma = \pi$.
(g) $\rho = 200\mu\textrm{m}^{-2}$.
(h) $\rho = 1000\mu\textrm{m}^{-2}$.
(i) Probability distribution of the angle $\Omega$ between the MTOC and the 
capture-shrinkage $\textrm{IS}_{1}$, $\gamma = \pi$.
       \label{fig:two_IS_combined_one_each_side}} 
 \end{figure}

In this section, we analyze the scenario when two IS employ different mechanisms.
The cortical sliding  has multiple advantages over the capture-shrinkage mechanism.
Given the radii of the whole IS and its center  $R_{\textrm{IS}}=2\mu\textrm{m}^{-2}$ and $R_{\textrm{CIS}}=0.4\mu\textrm{m}^{-2}$, respectively, the surface of the whole IS is $25\times$ larger than the surface of the IS center. 
Moreover, the cortical sliding dyneins attach on the whole MT, capture-shrinkage just at the end. 
Consequently, multiple filaments are attached to cortical-sliding dyneins during the entire simulation.
The capture-shrinkage dynein attach only when the tip of the MT intersects with the narrow center of the IS making 
the attachment of capture-shrinkage dyneins far less frequent.
All capture-shrinkage dyneins can be unattached for a long time.  
On the other hand, the capture-shrinkage mechanism has the advantage that the attached MTs 
form a narrow stalk 
assuring the alignment of dynein forces, as visualized in Fig. \ref{fig:sketch_two_IS_combined_one_each_side}d.
 \newline

The resulting repositioning process is shown in Movie S5 of the Supporting Materials and Methods, $\rho_{\textrm{IS}}^{1} = \tilde{\rho}_{\textrm{IS}}^{2} = 400 \mu\textrm{m}^{-2}$. 
The capture-shrinkage dyneins are located in the right $\textrm{IS}_{1}$.
The MTOC moves to the left IS, since the MTs attach immediately
to cortical sliding dyneins and the center of the right IS is not intersected by plus ends of MTs, visualized in Fig. \ref{fig:sketch_two_IS_combined_one_each_side}a. 
When the MTOC approaches the left IS, the cortical-sliding dyneins detach and, simultaneously, the tips of MTs passing through the center of the right IS attach to capture-shrinkage dyneins, visually demonstrated in 
Figs. \ref{fig:sketch_two_IS_combined_one_each_side}b and c. 
Since the capture-shrinkage mechanism is opposed by cortical sliding, MTs can detach from the capture-shrinkage dyneins. It takes several MTs to attach in the center of the IS at the same time to compete with the force of cortical sliding dyneins. As the force of capture-shrinkage dyneins outweighs the force of the cortical-sliding, the MTOC moves to the right center in the direction given by the MT stalk, visualized in Fig. 
\ref{fig:sketch_two_IS_combined_one_each_side}d. The capture-shrinkage dynein detach when the MTOC approaches the right IS.
Simultaneously, cortical sliding dyneins attach at the left IS,
visually demonstrated in Fig. \ref{fig:sketch_two_IS_combined_one_each_side}e, and the MTOC moves again to the left IS. 
 \newline

Fig. \ref{fig:two_IS_combined_one_each_side}a shows that
when $\gamma < \pi$
 the transition frequency steadily increases with the dynein density.
Fig. \ref{fig:two_IS_combined_one_each_side}b shows that when the densities are low, the average angle  between the MTOC
and the capture-shrinkage $\textrm{IS}_{1}$ 
 $\overline{\Omega}>>\frac{ \gamma}{2}$ indicating that the MTOC is predominantly located closer to the cortical sliding $\textrm{IS}_{2}$.
Moreover, the angle decreases with the increasing dynein density.
Average dwell times $\overline{T}_{d}^{2}$ close to cortical sliding $\textrm{IS}_{2}$
and  $\overline{T}_{d}^{1}$ close to capture-shrinkage $\textrm{IS}_{1}$
decrease and increase with increasing density, respectively,
see Fig. \ref{fig:two_IS_combined_one_each_side}c.
\newline

It can be seen in Figs. \ref{fig:two_IS_combined_one_each_side}d,e,g and h
that initially the MTOC travels to the cortical sliding $\textrm{IS}_{2}$ in all cases except one.
The MTOC travels to the capture-shrinkage $\textrm{IS}_{1}$  only in a highly improbable 
scenario when  plus ends of multiple MTs intersect the narrow IS center. 
When $\gamma = \frac{3\pi}{4}$ and $\rho = 200\mu\textrm{m}^{-2}$ the MTOC dwells in the proximity 
of the cortical sliding $\textrm{IS}_{2}$, see \ref{fig:two_IS_combined_one_each_side}d and f.
The transitions to the capture-shrinkage $\textrm{IS}_{1}$
are interrupted and the MTOC travels back to the cortical sliding $\textrm{IS}_{2}$(black).
When the MTOC finishes the transitions to the $\textrm{IS}_{1}$,
it dwells in its proximity for a short time and then returns to the 
$\textrm{IS}_{2}$(blue, red).
Multiple transitions to $\textrm{IS}_{1}$ rarely occur(green).
Interrupted transitions to $\textrm{IS}_{1}$ can be explained by  constantly attached cortical sliding dyneins overpowering the force of capture-shrinkage mechanism.
If the MTOC finishes transition to the $\textrm{IS}_{1}$, capture-shrinkage
dyneins detach and cortical sliding pulls the MTOC back to the $\textrm{IS}_{2}$.
To conclude, the cortical sliding has the dominance over the capture-shrinkage mechanism when $\rho_{\textrm{IS}}<600\mu\textrm{m}^{-2}$, since the MTOC is located predominantly closer to the $\textrm{IS}_{2}$, see Fig \ref{fig:two_IS_combined_one_each_side}b-d and f.
\newline

The Fig. \ref{fig:two_IS_combined_one_each_side}e shows that when $\rho = 1000\mu\textrm{m}^{-2}$,
the transitions towards the capture-shrinkage $\textrm{IS}_{1}$ are  mostly uninterrupted indicating that the capture-shrinkage mechanism
can compete with cortical sliding dyneins by capturing several MTs and forming MT stalk,
as visualized in Fig. \ref{fig:sketch_two_IS_combined_one_each_side}d.
Moreover, the MTOC dwells longer close to the capture-shrinkage
 $\textrm{IS}_{1}$, see Fig. \ref{fig:two_IS_combined_one_each_side}c, e and f,
 resulting in the decrease of the 
  average MTOC-$\textrm{IS}_{1}$ angle, see Fig. \ref{fig:two_IS_combined_one_each_side}b.
Therefore, capture-shrinkage mechanism gains the dominance over the cortical sliding mechanisms as the dynein density increases.
\newline

When  $\gamma < \pi$, the transition frequency increases with the dynein density, see Fig. \ref{fig:two_IS_combined_one_each_side}a, and therefore it increases as the capture-shrinkage mechanism becomes dominant.
The increasing density of capture-shrinkage dyneins increases the probability of dynein attachment and the formation of MTs stalk that can overcome the cortical sliding mechanism.
The formation of the MTs stalk results in complete transitions towards the capture-shrinkage $\textrm{IS}_{1}$ and in the steep decrease of
cortical sliding dwell times, see Figs.
\ref{fig:two_IS_combined_one_each_side}c and e.
However, the capture-shrinkage dwell times increase only slightly with the increasing density, see Fig. \ref{fig:two_IS_combined_one_each_side}c.
Regardless of dynein density, motors detach at the end of the transition and  
depolymerized MTs are unlikely to reattach, visually demonstrated 
in Fig. \ref{fig:two_IS_sketch}d.
Consequently, as the dynein density increases, capture-shrinkage mechanism becomes more able to pull the MTOC, but remains unable to hold it leading to the increased transition frequency.
\newline

 The case of $\gamma = \pi$ is unique since the transition frequency increases with the dynein density before reaching the peak at $\rho = 400\mu\textrm{m}^{-2}$ and then it slowly decreases,
see Fig. \ref{fig:two_IS_combined_one_each_side}a. 
The MTOC trajectories differ when considering multiple dynein densities. When $\rho = 200\mu\textrm{m}^{-2}$, the MTOC moves similarly to the case when $\gamma<\pi$.
The MTOC transitions
 to one IS, dwells there and then it moves to the second IS, see Fig. \ref{fig:two_IS_combined_one_each_side}g.
Fig. \ref{fig:two_IS_combined_one_each_side}i shows that the MTOC is predominantly located closer to the cortical sliding
IS when the dynein density is low.
When $\rho = 1000\mu\textrm{m}^{-2}$, the MTOC dwells in the proximity of the capture-shrinkage $\textrm{IS}_{1}$, see Fig. \ref{fig:two_IS_combined_one_each_side}i and, 
the transitions to the cortical-sliding $\textrm{IS}_{2}$ are infrequent and unfinished,
see Fig. \ref{fig:two_IS_combined_one_each_side}h.
When $\rho \geq 600\mu\textrm{m}^{-2}$, the dynein force is strong enough to pull the MTOC to the close proximity of the center of the capture-shrinkage 
$\textrm{IS}_{1}$,  Fig \ref{fig:two_IS_combined_one_each_side}i. 
In such a case
  almost all MTs are attached to the cortical sliding dynein at the distant $\textrm{IS}_{2}$, visually demonstrated in Fig \ref{fig:sketch_two_IS_combined_one_each_side}f. 
  The MTOC stays in the proximity of the capture-shrinkage $\textrm{IS}_{1}$, see Fig. \ref{fig:two_IS_combined_one_each_side}i, since the cortical-sliding dyneins pull the MTOC in different directions and oppose each other. Moreover, the MTOC is pulled back to the close IS by MTs occasionally  attached to capture-shrinkage dyneins in the center of the $\textrm{IS}_{1}$, visually depicted by the red short MT in Fig. \ref{fig:sketch_two_IS_combined_one_each_side}f. 
The Movie S6 of the Supporting Materials and Methods shows the process for $\gamma=\pi$ and $\rho_{\textrm{IS}}^1 = \tilde{\rho}^2_{\textrm{IS}}=1000\mu\textrm{m}^{-2}$. 
  \newline

The transition frequency decreases with the distance between the two IS when  $\gamma\leq\frac{2\pi}{3}$, see Fig.
\ref{fig:two_IS_combined_one_each_side}a, since the MTOC has to travel longer distance.
When $\gamma>\frac{2\pi}{3}$ the distance is compensated by the 
increased attachment probability in the center of the IS
caused by the increased probability density of MTs length corresponding to the circumferential distance between the two IS,
see Fig. \ref{fig:two_IS_sketch}f. 
Increased number of attached capture-shrinkage MTs leads to the decreased cortical sliding dwelling times as the $\gamma$ increases, see Fig. \ref{fig:two_IS_combined_one_each_side}.
The capture-shrinkage dwell times increase with $\gamma$, since the higher number of MTs pull the MTOC closer to the IS, see 
Figs. \ref{fig:two_IS_combined_one_each_side}f and i.

\subsubsection*{Combined mechanisms in both IS}

\begin{figure}[hbt!]
\centering
     \includegraphics[trim=0 280 0 0,clip,width=0.66\textwidth]{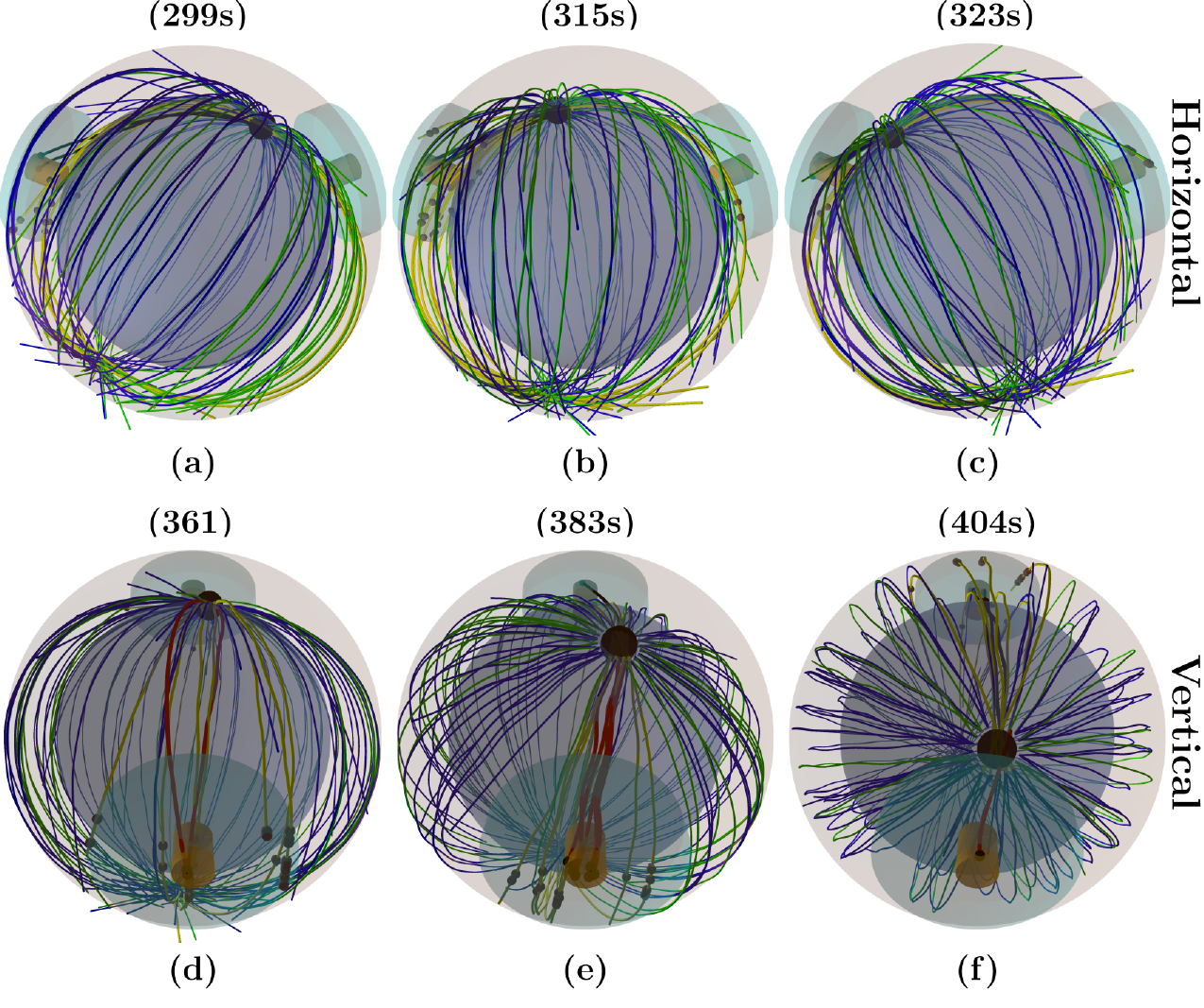}            
   \caption{Snapshots from the time-evolution of the MT cytoskeleton configuration under the effects of both mechanisms with the same density of the capture-shrinkage and cortical sliding dyneins in both IS $\tilde{\rho}_{\textrm{IS}}^{1}=\tilde{\rho}_{\textrm{IS}}^{2}=\rho_{\textrm{IS}}^{1}=\rho_{\textrm{IS}}^{2} = 400 \mu\textrm{m}^{-2}$
, $\gamma = \frac{3\pi}{4}$.
Brown and cyan cylinders indicate centers of both IS where capture-shrinkage dyneins are located and the whole areas of both IS containing cortical sliding dyneins, respectively.
Black and red lines represent MTs attached to capture-shrinkage dyneins in the centers of both IS, yellow lines depicts MTs attached to cortical sliding dyneins and blue and green lines indicate growing, shrinking unattached MTs, respectively. Small black spheres in both IS represent attached dyneins.
(a) The MTOC is closer to the right IS. MTs intersecting with the center of the IS attach to capture-shrinkage dyneins.
Cortical sliding dyneins attach to MTs at the periphery of the IS.
MTOC is pulled to the left IS by both mechanisms.  
(b-c) MTs captured in the left IS depolymerize as the MTOC approaches the IS.
Cortical sliding dyneins pull MTs at the periphery of the left IS.
MTs are not pulled by the capture-shrinkage dyneins from the left IS since no MT intersect with its center.
Cortical sliding dyneins at the right IS attach randomly on MTs, but they are 
overpowered by the combined force of both mechanism from the left IS and detach.
Consequently, substantially more attached cortical sliding dyneins pull at the left IS.
(d) As the MTOC approaches, MTs detach from the capture-shrinkage dyneins in the left IS.
Simultaneously, MTs intersect with the center of the right IS and are captured by dyneins.
Cortical sliding dyneins attach at the right IS and detach at the left.
(e) Both mechanism pull the MTOC to the right IS.
(f)As the MTOC approaches the right IS, dynein detach. Simultaneously, dynein attach in the distant IS.
\label{fig:sketch_two_IS_combined_both_equal}} 
 \end{figure}

The time-evolution of the cytoskeleton under the effect of both mechanisms with the equal densities in both IS $\tilde{\rho}_{\textrm{IS}}^{1}=\tilde{\rho}_{\textrm{IS}}^{2}=\rho_{\textrm{IS}}^{1}=\rho_{\textrm{IS}}^{2} = 400 \mu\textrm{m}^{-2}$, $\gamma = \frac{3\pi}{4}$, is shown in the Movie S7 of the Supporting Materials and Methods.
During the simulation, the MTOC repeatedly transitions between the two IS.
The snapshots of one transitions can be seen in  Fig. \ref{fig:sketch_two_IS_combined_both_equal}.
At the end of the transition, the MTs intersecting with the center of the distant IS are capture by dyneins, as visualized in \ref{fig:sketch_two_IS_combined_both_equal}a.
The cortical sliding dyneins in the right IS have to compete with both mechanisms from the left IS and detach.
Consequently, the MTOC is pulled to the left IS by both mechanisms and the movement is not opposed by the forces from the right IS, visualized in Figs. \ref{fig:sketch_two_IS_combined_both_equal}b and c.
As the MTOC approaches the left IS, capture-shrinkage MTs detach, visually in Fig.  \ref{fig:sketch_two_IS_combined_both_equal}d.
Simultaneously, MTs are captured on the other side of the cell.
Consequently, the stalk connecting the MTOC and the IS is formed and both mechanisms pull the MTOC to the right IS, as visualized in Fig.  \ref{fig:sketch_two_IS_combined_both_equal}e.
As the MTOC approaches the right IS dyneins of both mechanisms detach, visualized in \ref{fig:sketch_two_IS_combined_both_equal}f.
Simultaneously,  dyneins attach in the distant IS and they will initialize the next transition. 
\newline

\begin{figure}[hbt!]
\centering
      \includegraphics[trim=35 280 30 30,clip,width=0.66\textwidth]{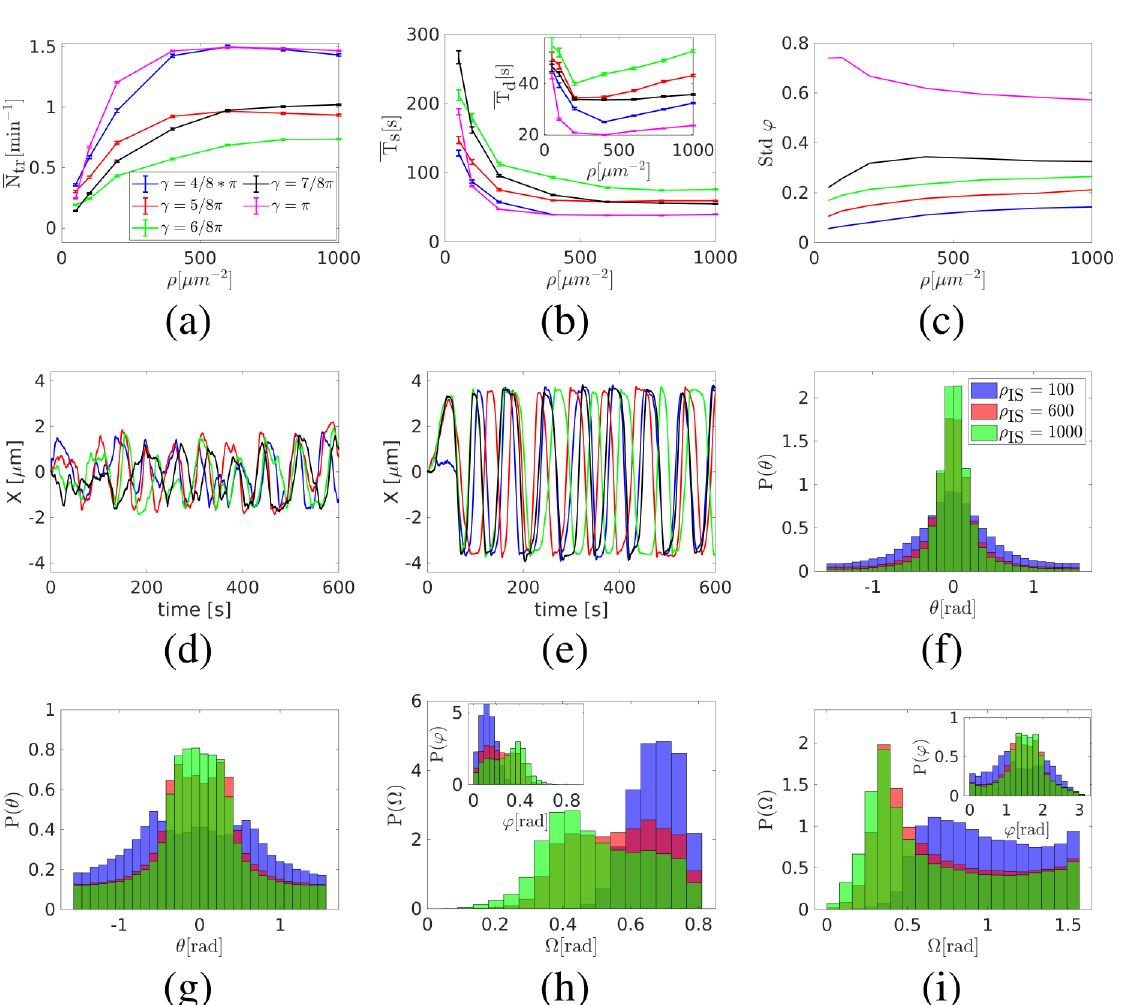}     
   \caption{Combination of capture-shrinkage and cortical sliding mechanisms with the same dynein density in both IS, 
$\tilde{\rho}_{\textrm{IS}}^{1}=\tilde{\rho}_{\textrm{IS}}^{2}=\rho_{\textrm{IS}}^{1}=\rho_{\textrm{IS}}^{2} = \rho$.
(a-c) The dependencies of the average transition frequency $\overline{N}_{\textrm{tr}}$ (a), the time that the MTOC spends in one hemisphere $\overline{T}_{s}$ and
 the dwell time in the proximity of the IS  $\overline{T}_{\textrm{d}}$(b) and the standard deviation of the polar angle $\varphi$ (c) on the dynein density $\rho$ are shown.  
(d and e) Examples of the time evolution of the MTOC position in 600s of simulation are shown.  The time evolutions of x coordinate of the MTOC are shown.
$\rho = 600\mu\textrm{m}^{-2}$ 
(d) $\gamma = \frac{\pi}{2}$. (e) $\gamma = \pi$.
(f-g) The probability distributions of the azimuthal angle $\theta$. (f) $\gamma = \frac{\pi}{2}$. (g)
 $\gamma = \pi$ (h-i) Probability distributions of the angle $\Omega$ between the MTOC and the $\textrm{IS}$  and the polar angle $\varphi$. (f) $\gamma = \frac{\pi}{2}$. (g)
 $\gamma = \pi$.
\label{fig:two_IS_combined_both_equal}} 
 \end{figure}

Fig. \ref{fig:two_IS_combined_both_equal}a shows that the transition frequency increases with the dynein density.
Moreover, the transition frequency decreases with the increasing angle $\gamma$ only when $\gamma\leq\frac{3\pi}{4}$
 and reaches the maximum when $\gamma=\pi$.
Surprisingly,  dwell times in the proximity of the IS do not steadily decrease with the dynein density
despite the continuous  decrease of the time that the MTOC spends in one hemisphere, see Fig. \ref{fig:two_IS_combined_both_equal}b.
The dwell times decrease with the dynein density until they reach a minimum when $\rho \sim 400 \mu\textrm{m}^{-2}$ and then they slightly increase. The standard deviation of the polar angle slightly increases and decreases when $\gamma<\pi$ and
$\gamma=\pi$, respectively.
\newline

By comparison between Figs. \ref{fig:two_IS_capture}a-c and \ref{fig:two_IS_combined_both_equal}d and e  one realizes that the MTOC trajectories under the effects of both mechanisms follow the same pattern as in the case of the sole capture-shrinkage mechanism: the MTOC travels to one IS, dwells in its close proximity and then transitions to the second IS. 
Moreover, the transitions between two IS are regular and continuous when $\gamma = \pi$ and incomplete and irregular when  $\gamma = \frac{\pi}{2}$.
As in the case of the capture-shrinkage mechanism, increasing circumferential distance between the two IS increases the 
probability that the plus end of a MT is captured in the center
of the distant IS due to the increasing probability density of MT length, see Fig. \ref{fig:two_IS_sketch}f.
Consequently, the dynein acts on increased number of filaments as the $\gamma$ increases assuring continuous transition.
\newline

The combination of mechanisms leads to the unprecedented transition frequency and shortest dwell times, compare
Figs. \ref{fig:two_IS_capture}, \ref{fig:two_IS_combined_one_each_side} and \ref{fig:two_IS_combined_both_equal}.
The reason is that
 the capture-shrinkage mechanism
supports the cortical sliding mechanism at the distant IS and 
hinders it at the close IS.
At the end of the transitions, 
capture-shrinkage MTs are depolymerized and the cortical sliding dyneins can attach to a lower number of MTs.
Contrarily, MTs attach to the capture-shrinkage dyneins in the distant center and the two mechanisms pull in alignment  sharing the load from opposing forces, as visualized in Figs. \ref{fig:sketch_two_IS_combined_both_equal}d and e.
Consequently, the MTOC is pulled to the distant IS by two mechanisms
 and to the close IS just by the cortical sliding acting on a reduced number of MTs.
\newline

The dwell times decrease with the rising dynein density,
see Fig. \ref{fig:two_IS_combined_both_equal}b, due to the higher pulling force.
The slight increase of dwelling times when 
$\rho>400\mu\textrm{m}^{-2}$ is caused by the fact the the MTOC travels closer to the IS, see Figs. \ref{fig:two_IS_combined_both_equal}h and i, and spends more time in the proximity of the IS. 
The monotonously  decreasing times that the MTOC spends in one hemisphere indicate that the process gets faster with the dynein density despite the slightly increased dwelling times, see 
Fig. \ref{fig:two_IS_combined_both_equal}b.
\newline

When $\gamma < \pi$, the standard deviation of the polar angle increases with the dynein density, see Fig. \ref{fig:two_IS_combined_both_equal}c, because the MTOC is pulled closer to the IS and the angle has a wider range, see Fig. \ref{fig:two_IS_combined_both_equal}h.
The standard deviation of the polar angle is the largest when $\gamma = \pi$, since the MTOC can transition between IS through the lower hemisphere, sketched in Fig. \ref{fig:two_IS_sketch}b. The standard deviation slightly decreases with the density.
The reason lies in the fast speed of the MTOC transitions that leads to the fast transition from one IS to the second.
Fig. shows \ref{fig:two_IS_combined_both_equal}i that
the MTOC is increasingly located closer to the IS when $\gamma=\pi$.
As in the case of the capture-shrinkage mechanism, the 
deviations from the $xz$ plane decrease with the rising density,
\ref{fig:two_IS_combined_both_equal}f and g.
The probability density does not have a peak at $\theta = 0$ when $\gamma = \pi$ at low densities, see Fig. \ref{fig:two_IS_combined_both_equal}g, 
 since the transitions pull the MTOC from the $xz$ plane and the force is often insufficient to finish the transition in the close proximity of the IS center.
\newline

\section*{Discussion}
We have analyzed the dynamics of the MTOC during the repositioning process
in situations a) in which the T cell has one IS in an arbitrary position with respect  to the initial position of the MTOC (quantifies by the angle 
$\beta$ sketched in Fig. \ref{fig:variable_Beta_basic}a ), and b) in which    the T cell has two IS at two arbitrary positions determined by the angle $\gamma$ between them, sketched in Fig. \ref{fig:two_IS_sketch}a.
In \cite{hornak_stochastic_2020}
 we studied the repositioning in the cell where the MTOC and the IS are initially diametrically opposed which was the configuration previously analyzed experimentally by Yi et al \cite{yi_centrosome_2013}.
 Here we showed that the predictions for this special situation are robust when more general, naturally occurring situations are considered. Most notably,
we found that the
 timescale for the completion of the relocation process agrees for a wide range of dynein densities. 
 Moreover, we predicted and provided explanations for the changes in the MT structure, the “opening” of the MT cytoskeleton resulting from the formation of the MT stalk and friction forces acting on unattached MTs. We further reported that the capture-shrinkage is the dominant mechanism in the cell when the MTOC and the IS are initially diametrically opposed. 
 We also discovered that the two mechanisms act in a fascinating synergy reducing the cell's resources for the efficient repositioning, since the combination of two mechanisms with a relatively low area density can be faster than the dominant mechanism with high density.
\newline

One of the differences occurring with smaller initial angles beta is that only a small fraction of MTs intersect the
IS when $\beta<\pi$, visually demonstrated in Fig. \ref{fig:variable_Beta_basic}d and Fig.
S1, leading to a smaller number of attached dyneins and slower repositioning,
see Fig. \ref{fig:variable_Beta}b.
Lower number of capture-shrinkage MTs and lower friction forces lead to milder changes in the cytoskeleton structure, compare Figs. 3f in \cite{hornak_stochastic_2020} and Fig. 
S3.
Nevertheless, the opening is still evident and can be used in experiments to prove the actions of capture-shrinkage mechanism regardless of cell's initial configuration.
\newline

It was shown in \cite{combs_recruitment_2006} that the dyneins colocalize with the ADAP ring in the pSMAC. Furthermore, it was hypothesized in 
\cite{kuhn_dynamic_2002} that one of the reasons that the PSMAC takes the form of the ring is to facilitate the interactions with MTs.
 Our work strongly supports such hypothesis, since the dynein attach to MTs predominantly at the IS periphery in all configurations \citep{hornak_stochastic_2020}, see Fig. S7 in
Supporting Materials and Methods.
 In our model, the cortical sliding dyneins were distributed homogeneously in the IS. However, attached dyneins are always located predominantly at the periphery of the IS. Moreover, attached dyneins move slightly to the periphery as the dynein area density increases, see Fig. S7a. 
 When the two mechanisms are acting together, the attached cortical sliding dyneins almost completely evacuate the center of the IS, see Fig. S9c in
Supporting Materials and Methods.
\newline

 The detailed study of cortical sliding repositioning revealed three different characteristics of the MTOC movement depending on three regimes for the dynein density. This behavior resulted from the competing forces of attached MTs sprouting from the MTOC in different directions. Such a behavior was not observed when $\beta<\pi$, see Fig. \ref{fig:variable_Beta} and Fig. S6, since attached MTs are aligned right from the beginning, as visualized in Fig. S5.
 \newline

 The comparison of the two mechanisms in various configurations demonstrated that the capture-shrinkage mechanism is dominant when $\beta>\pi/2$, since the times necessary for the MTOC repositioning are shorter, see Fig. \ref{fig:variable_Beta}b.
The cortical sliding mechanism is clearly dominant when $\beta<\pi/2$,  due to the absence of resisting force of the nucleus, see Fig. \ref{fig:variable_Beta}b. The case of $\beta=\pi/2$ appears as a borderline.
We can conclude that the two mechanisms have different roles in the cell.
When the initial positions of the MTOC and the IS are close to each other, the cortical sliding mechanism can assure effective repositioning.
Capture-shrinkage mechanism plays a key role when the cortical sliding reaches its limits so that a fast repositioning is assured in every configuration.
\newline

Most importantly, it was shown that the mechanisms act in synergy regardless the initial configuration of the cell, see Fig. 
S9a and b, since the dominant mechanism is always supported by the secondary one. The cortical sliding mechanism supports capture-shrinkage mechanism by passing MTs to it, see Figs. S9f and i.
The capture-shrinkage mechanism supports cortical sliding by providing a firm anchor point and pulling the MTOC from the nucleus, see Figs. S9g and h.
When the MTOC recedes from the nucleus, MTs copy the cell membrane more closely and the attachment to cortical sliding is more likely, see  Fig. S9e.
The dyneins of the two mechanisms pull in alignment sharing the load from opposing forces resulting in the decrease of the detachment probability.
The combination of two mechanisms with low area densities can be faster than the dominant mechanism with high densities \cite{hornak_stochastic_2020}, Fig. S9a and b. 
\newline

To conclude, the cell 
can polarize with a stunning efficiency by employing two mechanisms performing differently in various cell's configurations.
In the computational model the synergy of two mechanisms is displayed in the terms of speed. 
In the real cell where the cytoskeleton is dragged through 
the inhomogeneous environment of organelles and filaments, the synergy can make a difference between complete and unfinished repositioning.
Thus it appears that the location of dyneins on the IS periphery and the combination of two synergetically acting mechanisms together form a complex, efficient machinery assuring that the crucial immune response of human body is carried out efficiently while saving resources.
\newline

In situations in which the T cell has two IS (with relative positions defined by the angle $\gamma$, sketched in Fig. \ref{fig:two_IS_sketch}a) several scenarios have been observed experimentally \cite{kuhn_dynamic_2002} and are also predicted by our model: the MTOC alternates stochastically (but with a well defined average transition time) between the two IS; it wiggles in between the two IS without transiting to one of the two; or it is at some point pulled to one of the two IS and stays there. We have analyzed with the help of our model which scenario emerges in dependency of the mechanisms in action and the number of dyneins present. When
only the  capture-shrinkage mechanism is acting, the transition frequency increases and dwelling times decrease with increasing dynein density, see Figs. \ref{fig:two_IS_capture}d and e. 
Moreover, as the density increases, the lateral fluctuations of the MTOC (perpendicular to the $xz$ plane spanned by the centers of the two IS and the cell center) decrease for 
$\gamma\leq\frac{2\pi}{3}$, see Fig. \ref{fig:two_IS_capture_2_angle}a and b.
The increase of the angle $\gamma$ between the two IS changes the MTOC trajectories:
they are interrupted and incomplete when $\gamma=\frac{\pi}{2}$
and continuous when  $\gamma=\pi$.
One would expect that the transition frequency decreases with the distance between two IS.
Surprisingly, the transition frequency slightly decreases with increasing angle $\gamma$  only when $\gamma\leq\frac{2\pi}{3}$ and increases otherwise, see Fig. \ref{fig:two_IS_capture}.
The change of the MTOC trajectories and the increase of transition frequency with increasing $\gamma$ can be explained by the shape of the MT length distribution, see Fig. \ref{fig:two_IS_sketch}f.
As the $\gamma$ increases, increasing numbers of MTs have a length corresponding to the circumferential distance between two IS. 
Increasing numbers of attached MTs result in  stronger pulling force and a higher transition frequency.
Therefore, the presence of capture-shrinkage mechanism supports the transitions between two IS even when the densities are unequal, see Figs.
\ref{fig:two_IS_capture} and \ref{fig:two_IS_capture_2_unequal}. 
 \newline

When only the cortical sliding mechanism is present, the dyneins from both IS are in a constant tug-of-war. When the dynein densities are small, the MTOC wiggles around the central position, see Figs. \ref{fig:two_IS_cortical}d,g and j.
As the dynein density increases, one IS gains the upper hand and pulls the MTOC. 
Higher the density is, more the MTOC travels from the central position,
see Figs. \ref{fig:two_IS_cortical}f, i and l.
A subsequent transition to the distant IS is unlikely, since the dyneins from the distant IS have to overcome the forces from the close IS and 
the forces from the nucleus.
The effect  of the cortical sliding mechanism differs substantially from the effect of the capture-shrinkage mechanism, since the transition frequency decreases with increasing  angle $\gamma$ and with increasing  dynein density
 when $\tilde{\rho}_{\textrm{IS}}>200\mu\textrm{m}^{-2}$, see Fig. 
 \ref{fig:two_IS_cortical}a.
When $\gamma \geq \frac{3\pi}{4}$, the transition frequency is very small compared with the effect of the capture-shrinkage mechanism, compare Figs.  \ref{fig:two_IS_cortical}a
and \ref{fig:two_IS_capture}c.
In the special case of $\gamma=\pi$, the MTOC does not transition when 
 $\tilde{\rho}_{\textrm{IS}}\geq 600\mu\textrm{m}^{-2}$ due to the competing forces from the distant IS, sketched in Fig. \ref{fig:two_IS_sketch_cort_cort}c.
\newline

The mechanisms can be compared by locating them in different IS.
One observes that for $\rho_{\textrm{IS}}< 600\mu\textrm{m}^{-2}$, the average angle between the MTOC and the capture-shrinkage $\textrm{IS}_{\textrm{1}}$ $\overline{\Omega}>\frac{\gamma}{2}$, indicating that the MTOC is located closer to the cortical sliding $\textrm{IS}_{\textrm{2}}$, see Fig. \ref{fig:two_IS_combined_one_each_side}b.
As the density increases, the capture-shrinkage mechanism gains the upper hand and the MTOC is located closer to the capture-shrinkage $\textrm{IS}_{\textrm{1}}$, see Fig. \ref{fig:two_IS_combined_one_each_side}b, f and i.
One can therefore conclude that the cortical sliding mechanism is stronger only when $\rho_{\textrm{IS}}< 600\mu\textrm{m}^{-2}$ and weaker otherwise. 
The transition frequency increases with the dynein density when 
$\gamma<\pi$, see \ref{fig:two_IS_combined_one_each_side}a.
This is due to the fact that the capture-shrinkage dynein pull the MTOC towards the capture-shrinkage $\textrm{IS}_{\textrm{1}}$, the dyneins detach and the cortical sliding dyneins pull the MTOC to the distant IS.
When $\gamma=\pi$, the transition frequency decreases with increasing dynein density when $\rho_{\textrm{IS}}> 400\mu\textrm{m}^{-2}$.
In such a case, the MTOC moves closer to the IS as the dynein density increases,
see Fig. \ref{fig:sketch_two_IS_combined_one_each_side}i.
Subsequently, the MTOC dwells close to the capture-shrinkage IS because cortical sliding dyneins act against each other, as visualized  in
Fig. \ref{fig:sketch_two_IS_combined_one_each_side}f.
\newline

When the two mechanisms act together in both IS, the transition frequency increases with the dynein density, see Fig. \ref{fig:two_IS_combined_both_equal}a
and the dwell times are the lowest, see Fig. \ref{fig:two_IS_combined_both_equal}b.
The high transition frequency is due to the fact that as the MTOC is located in the proximity of one IS, the two mechanisms work together in the distant IS and oppose each other at the closer IS.
As the MTOC approaches the IS, the captured MTs depolymerize and at the end detach from dyneins. 
Consequently, the capture-shrinkage mechanism cannot keep the MTOC at the close IS and the cortical sliding mechanisms acts on a reduced numbers of MTs.
On the other hand, the dyneins from both mechanisms cooperate at the distant IS and share the load from opposing forces reducing their detachment rate.
 \newline

In conclusion we provided here a rather complete picture of the MTOC repositioning with one or two IS, under the model assumption of a fixed (spherical) cell shape. It would certainly be rewarding to include a deformable, semiflexible (due to the actin cortex) cell boundary interacting mechanically with the forces exerted by the semiflexible MTs. 
Another open question concerns the way in which dyneins are spatially organized in the membrane: do they self-organize \cite{hooikaas_kinesin-4_2020,gros_dynein_2021,sanchez_actin_2019} or are they more or less firmly anchored in the actin cortex as we assumed in our model. Probably more experimental insight is necessary to decide this question.

\section*{AUTHOR CONTRIBUTIONS}
I.H. and H.R. designed the research. I.H. performed calculations, prepared figures,
and analyzed the data. I.H. and H.R. wrote the manuscript.

\section*{ACKNOWLEDGMENTS}
This work was financially supported by the German Research Foundation
(DFG) within the Collaborative Research Center SFB 1027.

\section*{Declaration of Interest}
The authors declare no competing interests.


\section*{Supplementary Material}

An online supplement to this article can be found at
 \href{https://www.rieger.uni-saarland.de/homepage/research/biological_physics/research_publications/T_Cell_modelling.html}{
\textcolor{blue}{webpage of AG Rieger}}.

\end{document}